\documentclass[a4paper,11pt]{article}   
\usepackage{authblk} 
\usepackage[top=1.9cm,bottom=1.9cm,left=1.9cm,right=1.9cm]{geometry}

\usepackage{xspace}
\usepackage{tikz}
\usepackage{caption}
\usepackage{subcaption}
\usepackage{threeparttable}
\usetikzlibrary{mindmap}

\newcommand{\progname}{\textsf}
\newcommand{\compname}{\emph} 

\newcommand{\HTCondor}{\progname{HTCondor}\xspace}
\newcommand{\MW}{\progname{MW}\xspace}
\newcommand{\zram}{\progname{ZRAM}\xspace}
\newcommand{\tbb}{\progname{TBB}\xspace}
\newcommand{\bpp}{\progname{Bob++}\xspace}
\newcommand{\pth}{\progname{Pthread}\xspace}
\newcommand{\prs}{\progname{prs}\xspace}
\newcommand{\lrs}{\progname{lrs}\xspace}
\newcommand{\plrs}{\progname{plrs}\xspace}
\newcommand{\mplrs}{\progname{mplrs}\xspace}
\newcommand{\mts}{\progname{mts}\xspace}
\newcommand{\lrsa}{\progname{lrs1}\xspace}
\newcommand{\mplrsa}{\progname{mplrs1}\xspace}
\newcommand{\norm}{\progname{normaliz}\xspace}
\newcommand{\normaliz}{\progname{normaliz}\xspace}
\newcommand{\mai}{\compname{mai12}\xspace}
\newcommand{\mait}{\compname{mai20}\xspace}
\newcommand{\maicl}{\compname{mai}\xspace}
\newcommand{\maitt}{\compname{mai32}\xspace}
\newcommand{\mais}{\compname{mai64}\xspace}
\newcommand{\mainew}{\compname{mai32abcd}\xspace}
\newcommand{\maitf}{\compname{mai24}\xspace}
\newcommand{\maief}{\compname{mai32ef}\xspace}
\newcommand{\cdd}{\progname{cddr+}\xspace}
\newcommand{\porta}{\progname{PORTA}\xspace}
\newcommand{\ppl}{\progname{ppl\_lcdd}\xspace}
\newcommand{\lrslib}{\progname{lrslib}\xspace}
\newcommand{\tsubame}{\compname{Tsubame2.5}\xspace}
\newcommand{\gnuplot}{\progname{gnuplot}\xspace}

\newcommand{\polytope}{\emph}
\newcommand{\bvseven}{\polytope{bv7}\xspace}
\newcommand{\cthirty}{\polytope{c30}\xspace}
\newcommand{\cforty}{\polytope{c40}\xspace}
\newcommand{\fqfour}{\polytope{fq48}\xspace}

\newcommand{\mitseven}{\polytope{mit71}\xspace}
\newcommand{\mitine}{\polytope{mit}\xspace}
\newcommand{\permten}{\polytope{perm10}\xspace}
\newcommand{\cpsix}{\polytope{cp6}\xspace}
\newcommand{\vffive}{\polytope{vf500}\xspace}
\newcommand{\kmtwo}{\polytope{km22}\xspace}
\newcommand{\vfnine}{\polytope{vf900}\xspace}
\newcommand{\zfw}{\polytope{zfw91}\xspace}

\newcommand{\Adj}{\textrm{Adj}}
\newcommand{\initdepth}{\ensuremath{\mathit{init\_depth}}\xspace}
\newcommand{\mymaxdepth}{\ensuremath{\mathit{max\_depth}}\xspace}
\newcommand{\maxcobases}{\ensuremath{\mathit{max\_cobases}}\xspace}
\newcommand{\lmin}{\ensuremath{\mathit{lmin}}\xspace}
\newcommand{\lmax}{\ensuremath{\mathit{lmax}}\xspace}
\newcommand{\myscale}{\ensuremath{\mathit{scale}}\xspace}
\newcommand{\startvertex}{\ensuremath{\mathit{start\_vertex}}\xspace}
\newcommand{\mydepth}{\ensuremath{\mathit{depth}}\xspace}
\newcommand{\maxthreads}{\ensuremath{\mathit{max\_threads}}\xspace}
\newcommand{\numworkers}{\ensuremath{\mathit{num\_workers}}\xspace}
\newcommand{\mysize}{\ensuremath{\mathit{size}}\xspace}
\newcommand{\maxd}{\ensuremath{\mathit{maxd}}\xspace}
\newcommand{\maxc}{\ensuremath{\mathit{maxc}}\xspace}
\newcommand{\mystart}{\ensuremath{\mathit{start}}\xspace}

\newcommand{\unexplored}{\ensuremath{\mathit{unexplored}}\xspace}
\newcommand{\myfalse}{\ensuremath{\textrm{false}}\xspace}
\newcommand{\mytrue}{\ensuremath{\textrm{true}}\xspace}
\newcommand{\mycount}{\ensuremath{\mathit{count}}\xspace}
\newcommand{\putoutput}{\ensuremath{\textrm{put\_output}}\xspace}
\newcommand{\unfinished}{\ensuremath{\mathit{unfinished}}\xspace}

\usepackage{todonotes}
\usepackage{amsmath}
\usepackage{amssymb}
\usepackage{latexsym}
\usepackage{graphicx}
\usepackage{color}
\usepackage{hyperref}
\usepackage{slashbox}
\usepackage{algorithm}
\usepackage{algpseudocode}

\definecolor{darkblue}{rgb}{0,0,0.6}

\newcommand{\RR}{\mathbb{R}}

\begin{document}

\title{\mplrs: A scalable parallel vertex/facet enumeration code\thanks{This work was partially supported by JSPS 
Kakenhi Grants 16H02785, 23700019 and 15H00847, Grant-in-Aid for Scientific Research 
on Innovative Areas, `Exploring the Limits of Computation (ELC)'.}}

\author[1]{David Avis}
\author[2]{Charles Jordan}
\affil[1]{School of Informatics, Kyoto University, Kyoto, Japan and 
          School of Computer Science,
          McGill University, Montr{\'e}al, Qu{\'e}bec, Canada\\
          \texttt{avis@cs.mcgill.ca}}
\affil[2]{Graduate School of Information Science and Technology, 
          Hokkaido University, Japan\\
          \texttt{skip@ist.hokudai.ac.jp}}

\maketitle

\begin{abstract}
We describe a new parallel implementation, \mplrs, of the vertex enumeration code \lrs that 
uses the MPI parallel environment and can be run on a network of computers. 
The implementation makes use of a C wrapper that essentially uses the existing \lrs code
with only minor modifications. \mplrs was derived from the earlier parallel implementation \plrs,
written by G. Roumanis in C++ which runs on a shared memory machine.
By improving load balancing we are able to greatly improve performance for medium to large
scale parallelization of \lrs.
We report computational results comparing parallel and sequential codes for vertex/facet enumeration problems for convex polyhedra. 
The problems chosen span the range from simple to highly degenerate polytopes. 
For most problems tested, the results clearly show the advantage of using the parallel 
implementation \mplrs of the reverse search based code \lrs, even when as few as 8 cores are available. 
For some problems almost linear speedup was observed up to 1200 cores, the largest number of cores tested.

\noindent{}Keywords: vertex enumeration, reverse search, parallel processing\\
Mathematics Subject Classification (2000) 90C05
\end{abstract}

\section{Introduction}
\label{sec:intro}
The vast majority of mathematical programming software was designed and implemented for the prevalent
computers of the last century, which generally had single processors. Improvements in algorithmic design
and processor speed, in roughly equal measure, led to enormous speed-ups allowing increasingly
large problems to be solved in reasonable time. These legacy computer codes are sophisticated and extremely robust,
having been extensively tested on a wide range of platforms and applications. Previously,  
parallel processing was limited largely to expensive supercomputers. In recent years the situation
has changed radically;  increases in processor speed have been replaced by the ubiquitousness of multicore
processors. Desktop computers usually include at least four CPU cores and relatively inexpensive compute servers provide 
64 cores in a shared memory machine. Networks of such computers readily provide hundreds of available cores.
Unfortunately, very little legacy software for mathematical programming can make effective use of this hardware.

Some algorithms, such as those based on the simplex method, seem inherently sequential and will require new ideas
to exploit large scale parallel processing.
Others, such as integer programming via branch and cut, are basically tree searches that should benefit
greatly from parallelism. For example, the Concorde code for the travelling salesman problem~\cite{concorde,ABCC}
used large scale parallelism to solve extremely large problems to optimality over a distributed network.
General integer programming solvers such as CPLEX~\cite{cplex} and Gurobi~\cite{gurobi} also make use of multicore and distributed computing.
A computational study using Gurobi is contained in Koch et al.~\cite{Koch2012}. Using a shared memory machine, they report
speedups of roughly 9 times with 32 cores and 25 times with 128 cores for integer programming instances tested.
Using a distributed system with 8000 cores they report an estimated speedup of approximately 800 times.
However,
tests by Carle~\cite{Carle} show that results may be very disappointing if some processors are considerably slower than others, even with only 4 or 8 processors\footnote{See posts for December 9, 2014 (Gurobi) and
February 25, 2015 (CPLEX).}. Much work clearly needs to be done in this area.

In this paper we report on the parallelization of \lrs, a tree search algorithm for the {\em vertex/facet enumeration problem}.
The method we developed has the following features which are discussed in detail below:
(a) there is little modification to a complex legacy code;
(b) the parallelism is applied only in a wrapper;
(c) the subproblems are not interrupted;
(d) there is no communication between these threads;
and (e) it works on both shared-memory and distributed systems with essentially no user intervention required.
We also report computational results on a variety of problems and hardware that show near linear speedups, in some cases up to 1200 processors.

Vertex/facet enumeration problems find applications in many areas, of which
we list a few here\footnote{John White prepared a long list of applications
which is available at~\cite{lrs}.}.
Early examples include computing the facets of correlation/cut polyhedra by 
physicists (see, e.g.,~\cite{ceder1994,DL97}) and current research in 
this area relates to detecting quantum behaviour in computers such as D-Wave.
Research on facets of travelling salesman polytopes 
leads to important advances in branch-and-cut algorithms, see, e.g.,~\cite{ABCC}.
For example, Chv\'{a}tal local cuts are derived from
facets of small TSPs and this idea is also seen in the small instance relaxations
of Reinelt and Wenger~\cite{RW2003}.
Vertex enumeration is used to compute all Nash equilibria of bimatrix games and a code
for this based on \lrs is found at~\cite{lrs}. 
Vertex enumeration may be a last resort for minimizing extremely
complicated concave functions.
See, for example, Chapter 3 of Horst et al.~\cite{HPT02}.
This application shows the advantage of getting the output as a stream, most
of which can be immediately discarded.
When doing facet enumeration \lrs automatically computes the volume of the polytope
using much less memory than other methods, such as those described in \cite{FP16}.

The remainder of the paper is organised as follows.
We begin by introducing related work in Section~\ref{sec:relwork} and then proceed to
background on vertex enumeration, reverse search, \lrs and \plrs in Section~\ref{sect:back}.
This is followed in Section~\ref{sec:par} by a discussion of
various parallelization strategies that could be employed to manage the load balancing
problem. In Section~\ref{sec:mplrs} we discuss the implementation of \mplrs and
describe its features.
In Section~\ref{sec:experiments} we give some test results on a wide range of inputs,
comparing 7 codes: \cdd, \norm, \porta, \ppl, \lrs, \plrs and \mplrs and present an
analysis of our findings where we see that \mplrs scales further than other vertex
enumeration codes. Finally in the conclusion we compare our results with
those obtained for parallel integer programming and discuss the wider applicability of
our research.

\section{Related Work}
\label{sec:relwork}

We begin by reviewing the available algorithms and codes for vertex enumeration, focusing in particular on
parallel codes.  Codes for this problem were recently compared by Assarf et al.~\cite{polymake17},
however they focus primarily on sequential codes and therefore utilize comparatively
easy instances.  Then we introduce work on parallel reverse search, and also work on other parallel
search problems that may appear related to \mplrs.

There are basically two algorithmic approaches to the vertex enumeration problem: the Fourier-Motzkin double description method
(see, e.g.,~\cite{Ziegler})
and pivoting methods such as Avis-Fukuda reverse search~\cite{AF92} which enumerates all nodes of a tree.
The double description method involves inserting the half spaces from the H-representation sequentially and
updating the list of vertices that they span. 
Readily available codes for this method include \cdd~\cite{cdd},  
\norm~\cite{norm}, \ppl~\cite{ppl} and PORTA~\cite{porta}.
Although this sequential method did not seem easy to parallelize, it was recently achieved
and implemented in \norm. This breakthrough for the double description method involves
a new technique called pyramidal decomposition \cite{BIS16}. This decomposition is not equivalent to a standard
polyhedral decomposition and is much less costly to compute. We include experimental results
for \norm in Section \ref{sec:experiments}.

\subsection{\lrs and parallelization}

The reverse search method for vertex enumeration was implemented as \lrs~\cite{lrs,lrs2}.
From the outset it was realized that reverse search was eminently suitable for parallelization.
Marzetta developed the first parallel reverse search code using his \zram parallelization
platform~\cite{BMFN99,ZRAMthesis}, and implemented the first parallel vertex enumeration code, \prs, using
this generic reverse search framework. 
Load balancing is performed using a variant of what is now known as job stealing.
Application codes, such as \lrs, were embedded into \zram itself leading
to problems of maintenance as the underlying codes evolved.                     
Although \prs is no longer distributed and was based on a now obsolete version of \lrs,
it clearly showed the potential for large speedups of reverse search algorithms. 
Some limited experimental results for vertex enumeration are given in \cite{BMFN99} 
and these are discussed in Section \ref{parallel}.

The \lrs code is rather complex and has been under development for over twenty
years incorporating a multitude of different functions. It has been used
extensively and its basic functionality is very stable. Directly adding parallelization code
to such legacy software is extremely delicate and can easily produce bugs that
are difficult to find. A high level wrapper avoids this problem by implementing the parallelization
as a separate layer with very few changes to \lrs itself.
This allows the independent development of both parallelization
ideas and basic improvements in the underlying code, both of which stay up to date.
In return for this flexibility there are certain overheads that we discuss later.
However, the focus on \lrs and reverse search minimizes the number of modifications required
compared to using a general framework like \zram, and also allows the use of a load balancing technique
that is both simple and efficient for such codes.

The concept of a high level wrapper along these lines was tested by a shell
script, \progname{tlrs}, developed by White in 2009. Here the parallelization was achieved
by scheduling independent \lrs processes for subtrees via the shell. Although good speedups were obtained,
several limitations of this approach materialized as the number of processors available increased.
In particular job control becomes a major issue: there is no single controlling process.

To overcome these limitations the first author and Roumanis developed \plrs~\cite{AR13}.
This code is a C++ wrapper that compiles in the original \lrslib library essentially maintaining
the integrity of the underlying \lrs code. The parallelization was achieved by multithreading
using the Boost library and was designed to run on shared memory machines with little user interaction.
Experience with the \plrs
code showed good speedups with up to about 16 cores, then reduced
performance after that. 
The goal of \mplrs was to solve this load balancing problem and to move to a distributed
environment which could contain hundreds or thousands of processors.

The differences between \mplrs and \plrs are described in Section \ref{sec:plrs}.  While \prs
is able to run on distributed systems using the MPI layer in \zram, there are many differences
between \prs and \mplrs.  In particular, \prs uses a very different strategy for load balancing
where splitting work is distinct from performing work, splitting is computationally expensive, and
is targeted at cases with comparatively regular search trees (see Section 6.3.2 of \cite{ZRAMthesis}).
This is because the node descriptions are quite large (see Section 4.3.1 of \cite{ZRAMthesis}), and so
it tries to minimize the number of subproblems stored in memory.  \mplrs uses much smaller node descriptions (the cobasis)
and a very different strategy for load balancing, where splitting and performing work are not distinct.
This budgeted tree search results in the much better scaling and performance of \mplrs.
Many other differences between \prs and \mplrs are due to the age of \prs and 
the fact that it is no longer available or maintained.

\subsection{Other parallel codes}

The reverse search framework in \zram was also used to implement a parallel code for
certain quadratic maximization problems~\cite{FKL05}.
In a separate project, 
Weibel \cite{Weibel10} developed a parallel reverse search code to compute Minkowski sums.
This C++ implementation runs on shared memory machines and he obtains linear speedups with up to 8 processors,
the largest number reported.

\zram is a general-purpose framework that is able to handle a number of other applications, such
as branch-and-bound and backtracking, for which there are by now a large number of competing frameworks.
Recent papers by Crainic et al.~\cite{CLR06}, McCreesh et al.~\cite{MP15} and Herrera et al.~\cite{He17}
describe over a dozen such systems. While branch-and-bound may seem similar to reverse search enumeration, there
are fundamental differences. In enumeration it is required to explore the entire
tree whereas in branch-and-bound the goal is to explore as little of the tree as possible
until a desired node is found. The bounding step removes subtrees from consideration and
this step depends critically on what has already been discovered. Hence the order of
traversal is crucial and the number of nodes evaluated varies dramatically depending on this
order. Sharing of information is critical to the success of parallelization.
Similar complications exist in parallel SAT solving~\cite{HW12} and
parallel game tree search~\cite{HSN88}.
These issues do not occur in reverse search enumeration, and so a much lighter wrapper is possible.

Relevant to the heaviness of the wrapper and amount of programming effort required,
a comparison of three frameworks is given in \cite{He17}. The first, \bpp  \cite{Dj06},
is a high level abstract framework, similar in nature to \zram, on top of which the application sits.
This framework provides parallelization with relatively little programming effort
on the application side and can run on a distributed network. 
The second, Threading Building Blocks (\tbb) \cite{Re07},
is a lower level interface providing more control but also 
considerably more programming effort. It runs on a shared memory machine. 
The third framework is the \pth model \cite{Ca08} in which parallelization
is deep in the application layer and migration of threads is done by the operating system.
It also runs on a shared memory machine.
All of these methods use job stealing for load balancing \cite{BL99}. 
In \cite{He17} these three approaches are applied to a global optimization algorithm.
They are compared on a rather small
setup of 16 processors, perhaps due to the shared memory limitation of the last two approaches. 
The authors found that \bpp achieved a disappointing speedup of about 3 times, 
considerably slower than the other two approaches
which achieved near linear speedup. 
Other frameworks include CHiPPS~\cite{YanXu07} for parallel tree search
and \MW~\cite{Goux01}, which uses the \HTCondor framework.
\MW can be used to parallelize existing applications using the master-worker paradigm; one
such application was to quadratic assignment problems~\cite{ABGL02}.

Computational tasks that can be divided into subproblems which can be solved 
independently with no communication
are often called embarrassingly parallel~\cite{WA}.  Many such problems
involve processing an enormous amount of data that can easily be divided,
one prominent example being the SETI@home project~\cite{SETI}.
A recent approach to parallel constraint solvers~\cite{MRR16} (where the
input and output are comparatively small) uses this
as inspiration and initially creates a large number of 
subproblems that are then solved in parallel.  Other approaches to
creating an initial (hopefully balanced) decomposition of the input
include 
cube-and-conquer~\cite{HKWB11},
which uses a lookahead SAT solver to split the original problem into
many subproblems that are solved in parallel by CDCL solvers, and
applying machine learning techniques to parallel AND/OR 
branch-and-bound~\cite{OD17}.  Self-splitting~\cite{FMS14} is a technique
for minimizing communication when the subproblem descriptions are large.
There, each worker performs an identical split of the original problem and
then follows some deterministic rule to decide which portions belong to it.
This is not particularly appropriate in our case, where subproblem descriptions
are small and the major concern is that subproblem difficulty is
highly unbalanced.

Another way to deal with the problem posed by subproblems 
of varying difficulties is dynamic load balancing, where one
can split difficult subproblems during the computation.  Work
stealing~\cite{BL99} is one well-known approach where free workers
can steal portions of work from busy workers.

Parallel search has a long history and many applications~\cite{GK99}.
Topics related to this paper include load balancing techniques~\cite{KGR94,KR87} and estimating the difficulty of
subproblems.  The general idea of developing a lightweight parallel wrapper and
reusing sequential code with minimal changes has been applied in many areas,
including mixed integer programming~\cite{SABHK12} and SAT solving~\cite{BSS15}.

Parallel programming is almost as old as programming itself and there is a wealth of literature
on the subject which we can not cover here. For a modern introduction the reader is referred to Mattson et al.~\cite{MSM}.
Generally, much attention is given to machine architecture, communications between processes,
data sharing, synchronization, interrupts, load balancing and so on. This is essential knowledge for building and
implementing a parallel algorithm from scratch. However our aim was essentially different.
In return for some computational overhead, we would like to use existing sequential code with only
minor modifications. In particular, this eliminates the need for considering most of these topics.
The main issue that remains is load balancing, a topic we discuss in detail throughout the paper.

\section{Background}
\label{sect:back}
\subsection{The vertex/facet enumeration problem}
\label{sect:ve}
The vertex enumeration problem is described as follows.
Given an $m \times d$ matrix $A=(a_{ij})$ and an $m$ dimensional vector
$b$, a
\emph{convex polyhedron}, or simply \emph{polyhedron}, $P$ is defined as:
\begin{equation}
\label{polyhedron}
P=\{x\in \RR^d:b+Ax\geq 0\}.
\end{equation}
This description of a polyhedron is known as an \emph{H-representation}.
A \emph{polytope} is a bounded polyhedron. For simplicity in this article
we will assume that the input data $A,b$ defines 
a polytope which has dimension $d$, i.e.\ it is
full dimensional.
For this it is necessary that $m>d$. 
A point $x\in P$ is a \emph{vertex} of $P$ if and only if
it is the unique solution to a subset of $d$ inequalities from~(\ref{polyhedron})
solved as equations. Such a subset of inequalities is called a {\em basis}.

The \emph{vertex enumeration problem} is to output all vertices of a
polytope $P$. This list of vertices gives us a \emph{V-representation} of
$P$. The reverse transformation, called the \emph{facet enumeration problem},
takes a V-representation and computes its
H-representation. The two problems are computationally equivalent via polarity.
A polytope is called {\em simple} if each vertex is described by a single basis
and {\em simplicial} if each facet contains exactly $d$ vertices.
A vertex enumeration problem for a simple polytope is called {\em non-degenerate}
as is a facet enumeration problem for a simplicial polytope. Other
such problems are called {\em degenerate}.
Since the two problems are equivalent, we will consider only the vertex enumeration
problem in what follows.

One of the features of these types of enumeration problems is that the output
size varies widely for given
parameters $m$ and $d$. 
It is known that up to scaling by constants, each full dimensional polytope
has a unique  non-redundant $H$ and $V$ representation.
For the bounds given next we assume such representations.
For positive integers $m>d$ let
\def\lf{\left\lfloor}
\def\rf{\right\rfloor}
\begin{equation}
f(m,d)= \binom{m- \lf \frac{d+1}{2} \rf }{m-d}+ \binom{m- \lf \frac{d+2}{2} \rf }{m-d}\,.
\label{ubt}
\end{equation}
McMullen's Upper Bound Theorem (see, e.g.,~\cite{Ziegler})
states that for a polytope whose $H$-representation has parameters $m>d$ 
the maximum number of vertices it can have is $f(m,d)$. This bound is tight
and is achieved by the class of cyclic polytopes.
By inverting the formula and using polarity we can get lower bounds on the number of vertices of a polytope.
We have:
\begin{equation}
\min \{t: m \le f(t,d) \} ~\le ~|V| ~\le ~f(m,d)\,.
\label{bt}
\end{equation}
The first inequality follows because a polytope with fewer than this number of vertices must have less than $m$ facets.
For example, suppose $m=40$ and $d=20$. Then we have $22 \le |V| \le 40,060,020$. 

Pivoting methods compute the bases of a polytope and this number can be much larger than the upper bound in (\ref{bt}). 
However, as described in the next subsection, \lrs uses lexicographic pivoting which is equivalent to a symbolic perturbation
of the polytope into a simple polytope. Hence $f(m,d)$ is a tight upper bound on the number of
bases computed. Since we only require each vertex once, highly degenerate polytopes will
cause large overhead for pivoting methods. 

\subsection{Reverse search and \lrs}
\label{sec:lrs}

Reverse search is a technique for generating large, relatively unstructured, sets of discrete
objects. We give an outline of the method here and refer the reader to~\cite{AF92,AF93}
for further details.
In its most basic form, reverse search can be viewed as the traversal of a spanning tree, called the reverse
search tree $T$, of a graph $G=(V,E)$ whose nodes are the objects to be generated. Edges in the graph are
specified by an adjacency oracle, and the subset of edges of the reverse search tree are
determined by an auxiliary function, which can be thought of as a local search function $f$ for an
optimization problem defined on the set of objects to be generated. One vertex, $v^*$, is designated
as the \emph{target} vertex. For every other vertex $v\in V$,
repeated application of $f$ must generate a
path in $G$ from $v$ to $v^*$. The set of these paths defines the reverse search tree $T$, which has root $v^*$.

A reverse search is initiated at $v^*$, and only edges of the reverse search tree are traversed.
When a node is visited the corresponding object is output.  Since there is no possibility of
visiting a node by different paths, the nodes are not stored.  Backtracking can be performed in the
standard way using a stack, but this is not required as the local search function can be used for
this purpose. This implies two critical features that are essential for effective parallelization.
Firstly, it is not necessary to store more than one node of the tree at any
given time and no database is required for visited nodes. 
Secondly, it is possible to \emph{restart} the enumeration process from
any given node in the tree using only a description of this one node.
This contrasts with standard depth first search algorithms for which restart
is only possible with a complete database of visited nodes as well as the backtrack stack
to the root of the search tree.

In the basic setting described here a few properties are required. Firstly, the
underlying graph $G$ must be connected and an upper bound on the maximum vertex degree, $\Delta$, must
be known.  The performance of the method depends on $G$ having $\Delta$ as low as
possible.  The adjacency oracle must be capable of generating the adjacent vertices of some given
vertex $v$ sequentially and without repetition.  This is done by specifying a function  
$\Adj(v,j)$, where $v$ is a vertex of $G$ and $j = 1,2,\ldots,\Delta$.  Each value of $\Adj(v, j)$ is
either a vertex adjacent to $v$ or null. Each vertex adjacent to $v$ appears precisely once as $j$ ranges
over its possible values.  For each vertex $v \neq v^*$
the local search function $f(v)$ returns the tuple $(u,j)$ where $v = \Adj(u,j)$ such that $u$
is $v$'s parent in $T$.
Pseudocode is given in Algorithm~\ref{rsalg1}.
Note that the vertices are output as a continuous stream.
For convenience later, we do not output the root vertex $v^*$ in the pseudocode shown.

\begin{algorithm}
\begin{algorithmic}[1]
\Procedure{rs}{$v^*$, $\Delta$, $\Adj$, $f$}
        \State $v \gets v^*$~~~$j \gets 0$~~~$\mydepth \gets 0 $
        \Repeat
       	\While {$j < \Delta$}
                \State $j \gets j+1$
		\If {$f(\Adj(v,j)) = v$}  \Comment{forward step}
			\State $v \gets \Adj(v,j)~~~~~$  
			\State $j \gets 0$ 
                        \State $\mydepth \gets \mydepth+1$         
			\State {\bf output $v$}         
                \EndIf
        \EndWhile
        \If {$\mydepth > 0$}   \Comment{backtrack step}                 
		\State $(v,j) \gets f(v)$
                \State $\mydepth \gets \mydepth-1  $       
        \EndIf
        \Until {$\mydepth=0$ {\bf and} $j=\Delta$}
\EndProcedure
\end{algorithmic}
\caption{Generic Reverse Search}
\label{rsalg1}
\end{algorithm}

To apply reverse search to vertex enumeration we first
make use of \emph{dictionaries}, as is done for the
simplex method of linear programming.
To get a dictionary for (\ref{polyhedron}) we
add one new nonnegative variable for each inequality:
\[
x_{d+i}=b_i+\sum_{j=1}^{d}{a_{ij}x_j}, ~~
x_{d+i}\geq 0~~~  i=1,2,\ldots,m.
\]
These new variables are called \emph{slack variables} and the
original variables are called
\emph{decision variables}.

In order to have any vertex at all we must
have $m\geq d$, and normally $m$ is significantly larger than~$d$,
allowing us to solve the equations for various sets of variables on
the left hand side.
The variables on the left hand side of a dictionary are called
\emph{basic}, and those on the right hand side are called \emph{non-basic}
or, equivalently, \emph{co-basic}.
We use the notation
$B=\{i:{x_i}\mbox{ is basic}\}$ and
$N=\{j:{x_j}\mbox{ is co-basic}\}$.

A \emph{pivot} interchanges one index from $B$ and $N$ and solves the equations
for the new basic variables.
A \emph{basic solution} from a dictionary is obtained by setting
$x_j=0$ for all $j\in N$.
It is a \emph{basic feasible solution (BFS)} if $x_j\geq 0$ for every slack
variable $x_{j}$. A dictionary is called
\emph{degenerate} if it has a slack basic variable $x_j=0$.
As is well known, each BFS defines a vertex of $P$ and each vertex of $P$ can
be represented as one or more (in the case of degeneracy) BFSs. 

Next we define the relevant graph $G=(V,E)$ to be used in Algorithm~\ref{rsalg1}.
Each node in $V$ corresponds to a BFS and is labelled with the cobasic set $N$.
Each edge in $E$ corresponds to a pivot between two BFSs. 
Formally we may define the adjacency
oracle as follows. Let $B$ and $N$ be index sets for the current dictionary.
For $i \in B$ and $j \in N$
\[
\Adj(N,i,j) =
  \left\{ \begin{array}{ll}
N  \setminus \{j\} \cup \{i\} & \mbox{if this gives a feasible dictionary}\\
\emptyset & \mbox{otherwise\,.} \\
 \end{array}\right.
\]

A target $v^*$ for the
reverse search is found by solving a linear program over this dictionary
with any objective function. A new objective
function is then chosen so that
the optimum dictionary is unique and represents $v^*$.
\lrs uses Bland's least subscript rule for selecting the variable which
enters the basis and a lexicographic ratio test to select the leaving variable.
The lexicographic rule simulates a simple polytope which greatly reduces the number
of bases to be considered. We initiate the reverse search from the unique optimum dictionary.
For more details see the technical description at \cite{lrs}.
\lrs is an implementation of Algorithm~\ref{rsalg1} in exact rational arithmetic
using $\Adj, f,\mbox{ and }v^*$ as just described.

\subsection{Parallelization and \plrs}
\label{sec:plrs}

The development of \mplrs started from our experiences with \plrs, with
the goal of scaling past the limits of other vertex enumeration codes
while using the existing \lrs code with only minor modifications.
The details of \plrs are described in~\cite{AR13};
here we give a generic description of the parallelization
which is by nature somewhat oversimplified. We will use as an example the tree shown
in Figure~\ref{fig:mitfig} which shows the first two layers of the
reverse search tree for the problem \mitine, an 8-dimensional polytope with
729 facets that will be described in Section~\ref{subsec:expsetup}. 
The weight on each node is
the number of nodes in the subtree 
that it roots. The root of the tree is in the centre and its weight
shows that the tree contains 1375608 nodes, the number of cobases generated
by \lrs. At depth 2 there are 35 nodes but of these, just the four 
underlined nodes
contain collectively about 58\% of the total number of tree nodes.

\begin{figure}[htbp]
\centering
\begin{tikzpicture}[grow cyclic, align=flush center,
    level 1/.style={level distance=3cm,sibling angle=45},
    level 2/.style={level distance=1.6cm,sibling angle=30}]
\node{$1375608$}
 child { node {$197049$}
         child { node {$28441$} }
         child { node {$\underline{\textcolor{red}{148824} }$} }
         child { node {$5993$} }
         child { node {$13789$} }
         child { node {$1$} }
       }
 child { node {$33034$}
         child { node {$27911$} }
         child { node {$4346$} }
         child { node {$625$} }
         child { node {$151$} }
       }
 child { node {$328220$}
         child { node {$49456$} }
	 child { node {$7296$} }
	 child { node {$17443$} }
         child { node {$45263$} }
         child { node {$\underline{\textcolor{red}{208761}}$ } }
       }
 child { node {$439235$}
         child { node {$1555$} }
         child { node {$\underline{\textcolor{red}{308626}}$ } }
         child { node {$\underline{\textcolor{red}{129053}}$ } }
       }
 child { node {$141268$}
         child { node {$6963$} }
         child { node {$23914$} }
         child { node {$51832$} }
         child { node {$42903$} }
         child { node {$15591$} }
         child { node {$64$} }
       }
 child { node {$176210$}
         child { node {$26329$} }
         child { node {$22393$} }
         child { node {$45424$} }
	 child { node {$816$} }
         child { node {$81247$} }
       }
 child { node {$8$}
         child { node {$4$} }
         child { node {$3$} }
       }
 child { node {$60583$}
         child { node {$16123$} }
         child { node {$4678$} }
         child { node {$37747$} }
         child { node {$903$} }
         child { node {$1131$} }
       };
\end{tikzpicture}
\caption{Number of nodes in subtrees at depth 2 for \mitine}
\label{fig:mitfig}
\end{figure}
The method implemented in \plrs proceeds in three phases. 
In the first phase, sometimes called ramp-up in the parallel processing literature,
we generate the reverse search tree $T$ down to a fixed
depth, \initdepth, reporting all nodes to the output stream. In addition, the nodes of the tree 
with depth equal to  \initdepth which are not leaves of $T$
are stored in a list $L$. 

In the second phase we schedule subtree enumeration for nodes in $L$
using a user-specified parameter \maxthreads to limit the number of parallel processes. 
For subtree enumeration we use \lrs with a slight
modification to its earlier described restart feature.
Normally, in a restart, \lrs starts at a given restart node at its given 
depth and computes all remaining nodes in the tree $T$.
The simple modification is to supply a depth of zero with the restart node so that the search
terminates when trying to backtrack from this node. 

When the list $L$ becomes empty we move to Phase 3, sometimes called ramp-down,
in which the threads
terminate one by one until there are no more running and the procedure terminates.
In both Phase 2 and Phase 3 we make use of a \emph{collection process} which concatenates the output from the threads
into a single output stream.
It is clear that the only interaction between the parallel threads is the common output
collection process. The only signalling required is when a thread initiates or terminates
a subtree enumeration. 

\begin{figure}[htbp]
\centering
\includegraphics[width=0.8\textwidth]{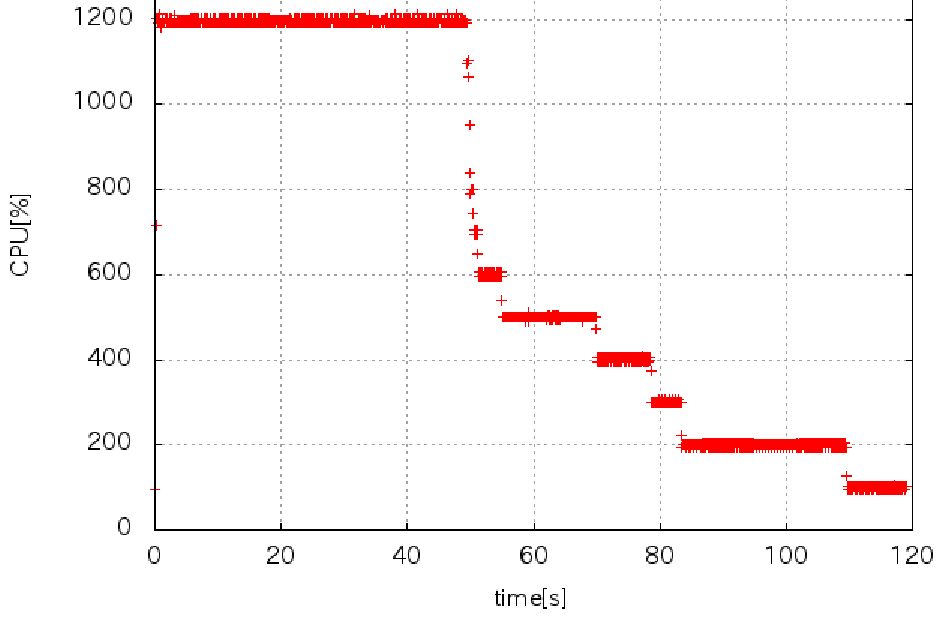}
\caption{Processor usage by \plrs on problem \mitine on a 12 core machine, $\initdepth=3$}                          
\label{fig:mit_L_12}
\end{figure}

Let us return to the example in Figure~\ref{fig:mitfig}.
Suppose we set $\initdepth=2$ and $\maxthreads=12$. 
A total of 35 nodes are found at this depth. 34 are stored in $L$ and the other, being a leaf, is ignored.
The first 12 nodes are removed from $L$ and scheduled on the 12 threads.
Each time a subtree is completely enumerated the associated thread receives
another node from $L$ and starts again. When $L$ is empty the thread is idle
until the entire job terminates.
To visualize the process refer to Figure~\ref{fig:mit_L_12}.
In this case we have set $\initdepth=3$ to obtain a larger $L$.
The vertical axis shows thread usage and the horizontal axis shows time.
Phase 1 is so short - less than one second - that it does not appear.
Phase 2 lasts about 50 seconds, when all 12 threads are busy.
Phase 3 lasts the remaining 70 seconds as more and more threads become idle.
If we add more cores, only Phase 2 will profit. Even with very many
cores the running time will not drop below 70 seconds and so this technique does not
scale well.
In comparing  Figures~\ref{fig:mitfig} and~\ref{fig:mit_L_12} we see that
the few large subtrees create an undesirably long Phase 3.
Going to a deeper initial depth helps to some extent, but this eventually 
creates an extremely long list $L$ with subsequent increase in overhead
(see~\cite{AR13} for more details). 
Nevertheless \plrs performs very well with up to about 32 parallel threads,
as we will see in Section~\ref{sec:experiments}.

In analyzing this method we observe that in Phase 1 there is 
no parallelization, in Phase 2 all available cores are used, and in Phase 3
the level of parallelization drops monotonically as threads terminate.
Looking at the overhead compared with \lrs we see that
this almost entirely consists of the amount of 
time required to restart the reverse
search process. In this case it requires the time to pivot the input matrix to a given 
cobasis, which is not negligible.
However a potentially greater cost occurs when $L$ is empty and threads are idle.
As the number of available processors increase this cost goes up, but the overhead of restarting
remains the same, for given fixed \initdepth.
This leads to conflicting issues in setting the 
critical \initdepth parameter. A larger value implies that: 
\begin{itemize}
\item
only a single thread is working for a longer
time,
\item
the list $L$ will typically be larger requiring more overhead in restarts but,
\item
the time spent in Phase 3 will typically be reduced.
\end{itemize}

The success in parallelization clearly depends on the structure of the tree $T$.
In the worst case it is a path and no parallelization occurs in Phase 2.
Therefore in the worst case we have no improvement  
over \lrs.
In the best case the tree is balanced so that the list $L$ can be
short reducing overhead and all threads terminate at more or less the same time.
Success therefore heavily depends on the structure of the underlying enumeration
problem.  

\section{Load Balancing Strategies}

\label{sec:par}
\plrs generates subproblems
in an initial phase 
based on a user supplied \initdepth parameter.  This tends to perform best
on balanced trees which, in practice, seem rather rare.
In \plrs, workers (except the initial Phase 1 worker) always
finish the subproblem that they are assigned.  However, there is
no guarantee that subproblems have similar sizes and as we have seen 
they can differ dramatically.  As we saw earlier, this can lead to a major loss of parallelism
after the queue $L$ becomes empty. 
Load balancing is the efficient distribution of work among a number of processors and is 
a well-studied area of parallel computation, see for example Shirazi et al.~\cite{SLB}.
The constraints of our parallelization approach described in the Introduction, such as no
interrupts or communication between subprocesses, greatly limits the methods available.
In this section we discuss various strategies we tried in developing \mplrs.
In particular, we focus~on:
\begin{itemize}
 \item estimating the size of subproblems to 
improve scheduling and create reasonably-sized problems,
 \item dynamic creation of subproblems, where we
can split subproblems at any time instead of only during the initial phase,
 \item using budgets for workers, who return after exploring
a budgeted number of nodes adding unfinished subproblems to $L$.
\end{itemize}

\subsection{Subtree Estimation}
\label{subsec:subtree}
A glance at Figure~\ref{fig:mitfig} shows the problem with
using a fixed initial depth to generate the subtrees for $L$:
the tree mass is concentrated on very few nodes. Of course,
increasing \initdepth would decrease the size of the large subtrees.
However, the subtrees can still be unbalanced at the new depth and
this also increases the number of jobs in 
$L$, increasing the restart overhead. Since \lrs has
the capability to estimate subtree size we tried two approaches using that: 
priority scheduling and iterative deepening.

Estimation is possible for
vertex enumeration by reverse
search using Hall-Knuth estimation~\cite{HK}.
From any node a child can be chosen at random and by continuing
in the same way a random path to a leaf is constructed. This leads
to an unbiased estimate of the subtree size from the initial node.
Various methods lead to lower variance, see~\cite{AD00}. 

The first use of estimation we tried was in priority scheduling.
Although finding a schedule that minimizes the total time to complete all work is NP-hard,
good heuristics are available.
One such heuristic is the list decreasing heuristic, analyzed by  Graham \cite{Gr69},
that schedules the jobs in decreasing order by their execution time.
Referring again to 
Figure~\ref{fig:mitfig}
we see that we should schedule those four heaviest subtrees at the start of
Phase 2. Since we do not have the exact values of the subtree sizes we
decided to use the estimation function as a proxy. We then scheduled jobs
from $L$ in a list decreasing manner by estimated tree~size.

A second idea we tried was iterative deepening.
We start by 
setting a threshold value, say $k$, for maximum estimated subtree size.  
Once a node at \initdepth is encountered an estimate of its subtree
size is made. If this exceeds $k$ then we continue to the next layer
of the tree and estimate the subtree sizes again, repeatedly going deeper
in the tree for subtrees whose estimates
exceed $k$. In this way all nodes returned to $L$ will have estimated
subtree sizes smaller than $k$.

The results from these two approaches were mixed. There are two negative
points. One is that Hall-Knuth estimates have very high variance, and
the true value tends to be larger than the estimate in probability.
So very large subtrees receiving small estimates would not be scheduled
first in priority scheduling and would not
be broken up by iterative deepening.
Secondly, the nodes visited during the random probes
represent overhead, as these nodes will all be visited again later.
In order to improve the quality of the estimate a large number of probes
need to be made, increasing this overhead.

Nevertheless this seems to be an interesting area of research. Newer
more reliable estimation techniques that
do not result in much overhead, such as the on-the-fly methods of~\cite{CKL}
and~\cite{KSW}, may greatly improve the effectiveness of these approaches.

\subsection{Dynamic Creation of Subproblems}
\label{subsec:dynamic}

As we saw in Section~\ref{sec:plrs}, \plrs creates new subproblems only
during the initial phase.  We can think in terms of one boss, 
who creates subproblems in Phase 1, and a set of workers who 
start work in Phase~2 and 
each works on a single subproblem until it is completed.
However, there is no reason why an individual worker cannot
send some parts of its search tree back to $L$ without exploring them.

A simple example of this is to implement a $\mathit{skip}$ parameter. 
This is set at some integer value $t>1$  and subtrees rooted
at every $t$-th node explored are 
sent back to $L$ without exploration. The boss
can set the $\mathit{skip}$ parameter dynamically when allocating work from $L$.
If $L$ is getting dangerously small, then a small value is set.
Conversely if $L$ is very large an extremely large value is set.

We implemented this idea but did not get good results. 
When the $\mathit{skip}$
parameter is set then all subtrees are split into smaller pieces,
even the small subtrees, which is undesirable.
When $\mathit{skip}$
is too small, the list $L$ quickly becomes unmanageably large
with very high overhead. 
It seemed hard for the boss to control the size of $L$
by varying the size of the parameter, due to the
delay incurred before the new parameter propagated to all the workers.

\subsection{Budgeted Subproblems}
\label{subsec:budg}

The final and 
most successful approach involved limiting
the amount of work a worker could do before being
required to quit. Each worker is given a \emph{budget}
which is the maximum number of nodes that can be visited.
Once this budget is exceeded the worker backtracks to the root
of its subtree returning all unfinished subproblems. These consist of all unexplored
children of nodes in the backtrack path.  This has several advantages.
Firstly, if the subtree has size less than the budget (typically 5000 nodes in practice)
then the entire subtree is evaluated without additional creation of overhead.
Secondly, each large subtree automatically gets split up. By including all
unexplored subtrees back to the root a variable number of jobs will be added to $L$.
A giant subtree will be split up many times. For example, the
subtree with 308626 nodes in Figure~\ref{fig:mitfig} will be split over 600 times, providing
work for idle workers. We can also change the budget dynamically to obtain different effects.
If the budget is set to be small we immediately create many new jobs for $L$. If
$L$ grows large we can increase the budget: since most subtrees will be below
the threshold the budget is not used up and  new jobs are not created. 

Budgeting can be introduced to the generic reverse search procedure of Algorithm~\ref{rsalg1}
as follows.
When calling the reverse search procedure we now supply three additional parameters: 
\begin{itemize}
\item
\startvertex is the vertex from which the reverse search should be initiated and replaces $v^*$,
\item
\mymaxdepth is the depth at which forward steps are terminated,
\item
\maxcobases is the number of nodes to generate before terminating and reporting unexplored subtrees.
\end{itemize}
Both \mymaxdepth and \maxcobases are assumed to be positive, for otherwise there is no work to do.
The modified algorithm is shown in Algorithm~\ref{brs}. 

\begin{algorithm}[htb]
\begin{algorithmic}[1]
\Procedure{brs}{\startvertex, $\Delta$, $\Adj$, $f$, \mymaxdepth, \maxcobases }
        \State $j \gets 0~~~v \gets \startvertex~~~\mycount \gets 0 ~~~\mydepth \gets 0$
        \Repeat
        \State $\unexplored \gets \myfalse$
        \While {$j < \Delta$ {\bf and} $\unexplored = \myfalse$ }
                \State $j \gets j+1$
                \If {$f(\Adj(v,j)) = v$}  \Comment{forward step}
                        \State $v \gets \Adj(v,j)~~~~~$
                        \State $j \gets 0$ 
                        \State $\mycount \gets \mycount+1$
                        \State $\mydepth \gets \mydepth + 1$
                        \If {$\mycount \ge \maxcobases$ {\bf or} $\mydepth = \mymaxdepth$} \Comment{budget is exhausted}
                            \State $\unexplored \gets \mytrue$
                        \EndIf
                        \State $\putoutput(v,\unexplored)$
                \EndIf
        \EndWhile
        \If {$\mydepth > 0$}   \Comment{backtrack step}
                \State $(v,j) \gets f(v)$
                \State $\mydepth \gets \mydepth - 1$
        \EndIf
        \Until {$\mydepth = 0$ {\bf and} $j=\Delta$}
\EndProcedure
\end{algorithmic}
\caption{Budgeted Reverse Search}
\label{brs}
\end{algorithm}
Comparing Algorithm~\ref{rsalg1} and Algorithm~\ref{brs} we note several changes. 
Firstly, an integer variable \mycount is introduced to keep track
of how many tree nodes have been generated. 
Secondly, a flag \unexplored is introduced to distinguish the tree nodes which have
not been explored and which are to be placed on $L$. It is initialized as \myfalse on line 4.
The flag is set to \mytrue in line 13 if either 
the budget of \maxcobases has been exhausted
or a depth of \mymaxdepth has been reached. 
In any case, each node encountered on a forward step is output via the routine \putoutput
on line 15.
In single-processor mode the output is simply sent to the output file with a flag added
to unexplored nodes. In multi-processor mode, the output is synchronized and
unexplored nodes are returned to $L$ (cf. Section~\ref{sec:mplrs}).

Backtracking is as in Algorithm~\ref{rsalg1}.
After each backtrack step the \unexplored flag is set to \myfalse in line 4.
If the budget constraint has been exhausted then \unexplored will again be set
to \mytrue in line 13 after the first forward step.
In this way all unexplored siblings of nodes on the backtrack path to the root are flagged
and placed on $L$. If the budget is not yet exhausted, forward steps continue until the budget is exhausted, \mymaxdepth is reached, or we reach a leaf.

To output all nodes in the subtree of $T$ rooted at $v$ we set  
$\startvertex=v$, $\maxcobases=+\infty$ and $\mymaxdepth=+\infty$.
So if $v=v^*$ this reduces to Algorithm~\ref{rsalg1}.
For budgeted subtree enumeration we set \maxcobases to be the worker's budget.
To initialize the parallelization process we will generate the tree $T$ down to a 
a small fixed depth with a small budget constraint in order to generate a
lot of subtrees. We then increase the budget constraint 
and remove the depth constraint so that most workers will finish the tree they
are assigned without returning any new subproblems for $L$.
Since subproblems are dynamically created, it is not necessary to have a long
Phase 1.  By default, \mplrs logs the time spent in Phase 1 and this time was
insignificant in all runs considered in this paper.
The details are given in Section~\ref{subsec:mplrs_master}.

\section{Implementation of \mplrs}
\label{sec:mplrs}

The primary goals of \mplrs were to move beyond single, shared-memory
systems to clusters and improve load balancing when a large number of cores is available.
The implementation uses MPI, and starts a user-specified number of
processes on the cluster.  One of these processes becomes the \emph{master},
another becomes the \emph{consumer}, and the remaining processes are
\emph{workers}.

The master process is responsible for distributing the input file and
parametrized subproblems to the workers, informing the other processes to exit
at the appropriate time, and handling checkpointing.  The consumer
receives output from the workers and produces the output file.  The workers
receive parametrized subproblems from the master, run the \lrs
code,  send output to the consumer, and return unfinished subproblems 
to the master if the budget has expired.

\subsection{Master Process}
\label{subsec:mplrs_master}

The master process begins by sending the input to all workers,
which may not have a shared file system.
In \mplrs, $\Adj$ and $f$ are defined as in
Section~\ref{sec:lrs} and so it suffices to send the input polyhedron.
Pseudocode for the master is given
in Algorithm~\ref{alg:mplrs_master}.

Since we begin from a single \startvertex, the master chooses an
initial worker and sends it the initial subproblem.  We cannot yet
proceed in parallel, so the master uses user-specified (or
very small default) 
initial parameters
\initdepth and \maxcobases to ensure that this worker will
return (hopefully many) unfinished subproblems quickly.
The master then executes its main loop, which it continues until
no workers are running and the master has no unfinished subproblems.
Once the main loop ends, the master informs all processes to finish.
The main loop performs the following tasks:
\begin{itemize}
 \item if there is a free worker and the master has a subproblem, subproblems
       are sent to workers;
 \item we check if any workers are finished, mark them as free and receive 
       their unfinished subproblems.
\end{itemize}
\begin{algorithm}[htb]
 \caption{Master process}
 \label{alg:mplrs_master}
 \begin{algorithmic}[1]
  \Procedure{mprs}{\startvertex, $\Delta$, $\Adj$, $f$,
   \initdepth, \mymaxdepth, \maxcobases, \lmin, \lmax, \myscale, \numworkers}
   \State {\bf Send} ($\Delta$, $\Adj$, $f$) to each worker
   \State {\bf Create empty list} $L$
   \State $\mysize \gets \numworkers + 2$
   \State {\bf Send} (\startvertex, \initdepth,
                      \maxcobases) to worker $1$
   \State {\bf Mark} $1$ as working

   \While {$L$ is not empty or some worker is marked as working}
          \While {$L$ is not empty and some worker not marked as working}
             \If {$|L|<\mysize\cdot \lmin$} 
                \State $\maxd \gets \mymaxdepth$
             \Else \State $\maxd \gets \infty$
             \EndIf
             \If {$|L|>\mysize\cdot \lmax$} 
                \State $\maxc \gets \myscale\cdot \maxcobases$
             \Else \State $\maxc \gets \maxcobases$
             \EndIf
          \State {\bf Remove} next element \mystart from $L$
          \State {\bf Send} (\mystart, \maxd, \maxc)
                            to first free worker $i$
          \State {\bf Mark} $i$ as working
          \EndWhile

          \For {each marked worker $i$}
            \State {\bf Check} for new message \unfinished from $i$
            \If {incoming message \unfinished from $i$}
               \State{\bf Join} list \unfinished to $L$
               \State{\bf Unmark} $i$ as working
            \EndIf
          \EndFor
     \EndWhile
     \State {Send} \texttt{terminate} to all processes
  \EndProcedure
 \end{algorithmic}
\end{algorithm}

Using reasonable parameters is critical to achieving good parallelization.
As described in Section~\ref{subsec:budg}, this is done dynamically
by observing the size of $L$. We use the parameters \lmin, \lmax and \myscale.
Initially, to create a reasonable size list $L$, we set  $\mymaxdepth=2$ and $\maxcobases=50$.
Therefore the initial worker will generate subtrees at depth 2 until 50 nodes have
been visited and then backtrack. Additional workers are given the same aggressive parameters
until $L$ grows larger than \lmax times the number of processors. We now multiply
the budget by \myscale and remove the \mymaxdepth constraint. Currently $\myscale=100$ so
workers will
not generate any new subproblems unless their tree has at least 5000 nodes.
If the length of $L$ drops below this bound we return to the earlier value of
$\maxcobases=50$ and if it drops below \lmin times the size of $L$ we
reinstate the \mymaxdepth constraint. The current default is to set $\lmin=\lmax=3$.
In Section~\ref{subsec:histo} we show an example of how the length of $L$
typically behaves with these parameter settings. 

\subsection{Workers}

The worker processes are simpler -- they receive the problem at
startup, and then repeat their main loop: receive a
parametrized subproblem from the master, work on it subject to the
parameters, send the output to the consumer, and send unfinished
subproblems to the master if the budget is exhausted.  

\begin{algorithm}[ht!]
 \caption{Worker process}
 \label{alg:mplrs_worker}
 \begin{algorithmic}[1]
  \Procedure{worker}{}
   \State{\bf Receive} ($\Delta$, $\Adj$, $f$) from master
   \While {\mytrue}
     \State{Wait for} message from master
     \If {message is \texttt{terminate}}
       \State {\bf Exit}
     \EndIf
     \State{\bf Receive} (\startvertex, \mymaxdepth, \maxcobases) 
     \State{\bf Call} BRS(\startvertex, $\Delta$, $\Adj$, $f$, \mymaxdepth,  \maxcobases)
     \State{\bf Send} list of unfinished vertices to master
     \State{\bf Send} output list to consumer
   \EndWhile
  \EndProcedure
 \end{algorithmic}
\end{algorithm}

\subsection{Consumer Process}

The consumer process in \mplrs is the simplest.  The workers send output
to the consumer in exactly the format it should be output (i.e., this
formatting is done in parallel).  The consumer simply sends it to an
output file, or prints it if desired.  By synchronizing output to a
single destination, the consumer delivers a continuous output
stream to the user in the same way as 
\lrs does.

\begin{algorithm}
 \caption{Consumer process}
 \label{alg:mplrs_consumer}
 \begin{algorithmic}[1]
  \Procedure{consumer}{}
   \While {\mytrue}
    \State{\bf Wait} for incoming message
    \If {message is \texttt{terminate}}
     \State {\bf Exit}
    \EndIf
    \State{\bf Output} this message
   \EndWhile
  \EndProcedure
 \end{algorithmic}
\end{algorithm}

\subsection{Histograms}
\label{subsec:histo}

There are additional features supported by \mplrs that
are minor additions to 
Algorithms~\ref{alg:mplrs_master}--\ref{alg:mplrs_consumer}.
We introduce \emph{histograms} in this subsection, before proceeding
to checkpoints in Section~\ref{subsec:checkp}.

When desired, \mplrs can provide a variety of information in a
histogram file.  Periodically, the master process
adds a line to this file, containing the following information:
\begin{itemize}
 \item real time in seconds since execution began,
 \item the number of workers marked as working,
 \item the current size of $L$ (number of subproblems the master has).
\end{itemize}

\begin{figure}[h!tb]
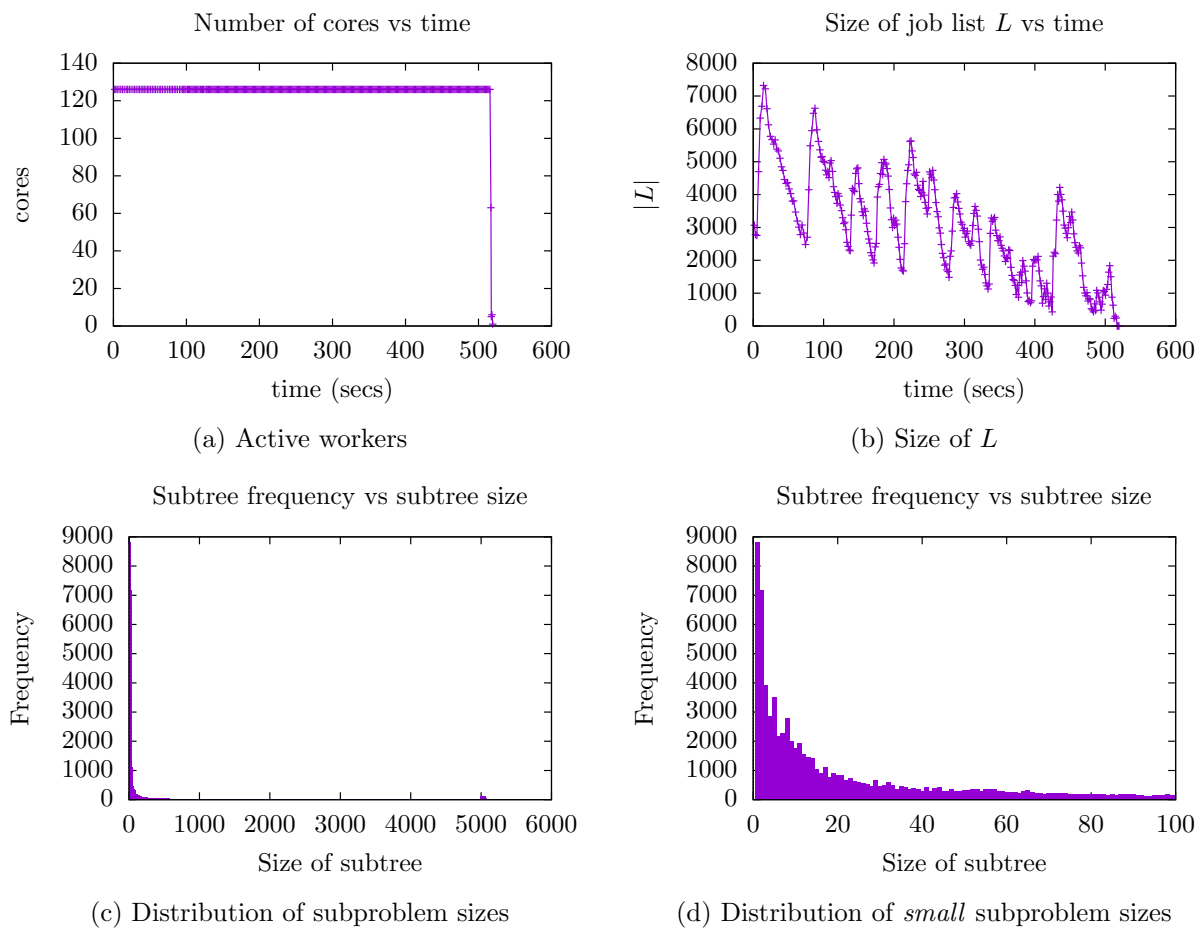

\centering
\begin{subfigure}[b]{0.47\textwidth}
 \centering
 \resizebox{\textwidth}{!}{
  \input{plots/hist-mit71_128-activeworkers.tex}
 }
 \caption{Active workers}
 \label{subfig:actwork}
\end{subfigure}
\begin{subfigure}[b]{0.47\textwidth}
 \centering
 \resizebox{\textwidth}{!}{
  \input{plots/hist-mit71_128-sizeL.tex}
 }
 \caption{Size of $L$}
 \label{subfig:size_L}
\end{subfigure}
\begin{subfigure}[b]{0.47\textwidth}
 \centering
 \resizebox{\textwidth}{!}{
  \input{plots/plotD-mit71_128.tex}
 }
 \caption{Distribution of subproblem sizes}
 \label{subfig:plotD}
\end{subfigure}
\begin{subfigure}[b]{0.47\textwidth}
 \centering
 \resizebox{\textwidth}{!}{
  \input{plots/plotD-mit71_128-small.tex}
 }
 \caption{Distribution of \emph{small} subproblem sizes}
\label{subfig:plotDa}
\end{subfigure}
\caption{Histograms for \mitseven with 128 processes}
\label{fig:hist_sample}
\end{figure}

We use this histogram file with \gnuplot to produce
plots that help understand how much parallelization is achieved over
time, which helps when tuning parameters.  Examples of the resulting output
are shown in Figure~\ref{fig:hist_sample}. The problem, \mitseven, 
is a degenerate 60-dimensional polytope with 71 facets
and is described in
Section \ref{subsec:expsetup}. 

It is useful to compare Figure~\ref{fig:hist_sample}(a)
to Figure \ref{fig:mit_L_12}
showing a typical \plrs run. The long Phase 3 ramp-down
time of \plrs no longer appears. This is due to the budget constraint
automatically breaking up large subtrees and the master redistributing this 
new work to other workers.  The fact that workers are generally not idle  
is necessary for efficient parallelization, but it is
not sufficient: if the job queue is very large the overhead required to start
jobs will dominate and performance is lost. To get information on
this the second histogram,   
Figure~\ref{fig:hist_sample}(b), is of use.
This plot gives the size of $L$, the number of subproblems held by the
master. This histogram is useful to visualize the overall progress of the run
in real time
to see if the parameters are reasonable.
In \mplrs, $L$ is implemented as a stack.
When $|L|$ falls to a value for the first time, a new (relatively high in
the tree) subproblem is examined for the first time.  If this new subproblem
happens to be large, the size of $L$ can grow dramatically due
to the budget being exhausted by the assigned worker.  The choice of
parameters greatly affects the rate at which new subproblems are created.

A third type of histogram, subtree size, can also be produced as shown
in Figure~\ref{fig:hist_sample}(c). 
This gives the frequency of the sizes of all subtrees whose roots were stored in
the list $L$, which in this case contained a total of 116,491 subtree roots.
We see that for this problem the vast
majority of subtrees are extremely small. The detail of this is
shown in Figure~\ref{fig:hist_sample}(d). These small subtrees
could have been enumerated more quickly than their restart cost alone
 -- if they could have been identified quickly.
This is an interesting research problem.  After about 60 nodes the
distribution is quite flat until the small hump occurring at 5000 nodes.
This is due to the budget limit of 5000 causing a worker to terminate.
The hump continues slightly past 5000 nodes reflecting the additional
nodes the worker visits on the backtrack path back to the root.
It is interesting that most workers completely finish their
subtrees and only very few actually hit the budget constraint.
Histograms such as these may be of interest for
theoretical analysis of the budgeting method. For example, the shape of
the histogram may suggest an appropriate random tree model to study
for this type of problem.

\subsection{Checkpointing}
\label{subsec:checkp}

An important feature of \mplrs is the ability to
checkpoint and restart execution with potentially different
parameters or number of processes.  This allows, for example, users
to tune parameters over time using the histogram file, without discarding
initial results. It is also very useful for very large jobs
if machines need to be
turned off for any reason or if new machines become available.

Checkpointing is easy to implement in \mplrs but to be effective
it depends heavily on the \maxcobases option being set.  
Workers are never aware of
checkpointing or restarting -- as in Algorithm~\ref{alg:mplrs_worker} they 
simply use \lrs to solve given subproblems until their budget runs out.
When the master wishes to checkpoint, it ceases distribution of new subproblems
and tells workers to terminate.  Once all workers have finished and returned
any unfinished subproblems, the master informs the consumer of a checkpoint.
The consumer then sends various counting statistics to the master, which
saves these statistics and $L$ in a \emph{checkpoint file}.
Note that if \maxcobases is not set then each worker must completely
finish the subtree assigned, which may take a very long time.

When restarting from a checkpoint file, the master reloads $L$ from the
file instead of
distributing the initial subproblem.  It informs the consumer of the counting
statistics and then proceeds normally.  Previous output is not re-examined: 
\mplrs assumes that the checkpoint file is correct.

\section{Performance}
\label{sec:experiments}

We describe here some experimental results for the
three codes described in this paper and 4 codes based on the double description method:
\cdd~\cite{cdd}, \norm~\cite{norm}, \porta~\cite{porta} and \ppl~\cite{ppl}.

\subsection{Experimental Setup}
\label{subsec:expsetup}

The tests were performed using the following computers:
\begin{itemize}
\item
\mait: 2x Xeon E5-2690v2 (10-core 3.0GHz), 20 cores, 128GB memory, 3TB hard drive,
\item
\maief: 4x Opteron 6376 (16-core 2.3GHz), 64 cores, 256GB memory, 4TB hard drive,
\item
\mainew: 4 nodes, each containing: 2x Opteron 6376 (16-core 2.3GHz), 32GB memory, 500GB hard drive (128 cores in total),
\item
\mais: 4x Opteron 6272 (16-core 2.1GHz), 64 cores, 64GB memory, 500GB hard drive,
\item
\mai: 2x Xeon X5650 (6-core 2.66GHz), 12 cores, 24GB memory, 60GB hard drive,
\item
\maitf: 2x Opteron 6238 (12-core 2.6GHz), 24 cores, 16GB memory, 600GB RAID5 array,
\item
\tsubame: supercomputer located at Tokyo Institute of Technology, nodes containing: 2x Xeon X5670 (6-core 2.93GHz), 12 cores, 54GB memory, large file systems, dual-rail QDR Infiniband.
\end{itemize}
The first six machines total 312 cores, are located at Kyoto University and connected with gigabit ethernet. They were purchased between 2011-15 for a combined
total of 3.9 million yen (\$33,200).

The polytopes we tested are described in Table \ref{polytopes} and range from non-degenerate
to highly degenerate polyhedra. 
The input for a vertex enumeration problem, as defined in (\ref{polyhedron}),
is given as an $m$ by $n$ array of integers or rationals, where $n=d+1$.
For $i=1,\ldots,m$, row $i$ consists of $b_i$ followed by the $d$ coefficients of the $i$-th row of $A$.
For a $d$-dimensional facet enumeration problem, $m$ is the number of vertices.
Each row has $n=d+1$ columns each consisting of a 1 (a 0 would represent an extreme ray)
followed by the $d$ coordinates of the vertex.
Table \ref{polytopes} includes the results of an \lrs run on each
polytope as  \lrs
gives the number of bases in a symbolic perturbation of the polytope.
We include a column labelled degeneracy which is the number of bases divided by the number of vertices (or facets) output, rounded to the
nearest integer. We have sorted the table in order of increasing degeneracy. The horizontal line separates the
non-degenerate from the degenerate problems.
The corresponding input files are available by following the {\em Download}
link  at~\cite{lrs}.
Note that the input sizes are small, roughly comparable and except for \cpsix, much smaller than the output
sizes.
Five of the problems were previously used in \cite{AR13}:
\begin{itemize}
\item \cthirty, \cforty: cyclic polytopes which achieve the upper bound (\ref{ubt}).
These have very large integer coefficients, the longest having 23 digits
for \cthirty and 33 digits for \cforty. The polytopes are given by their V-representation. Due to the internal lifting performed
by \lrs these appear to have degeneracy less than 1, but they are in fact non-degenerate
simplicial polyhedra.
\item \permten: the permutahedron for permutations of length 10, whose vertices are
the $10!$ permutations of $(1,2,3,\ldots,10)$. It is a 9-dimensional simple
polytope. More generally, for permutations of length $p$, this polytope is described by $2^p -2$ facets and one equation
and has $p!$ vertices.
The variables all have coefficients $0$ or $1$.
\item \mitine: a configuration polytope used in materials science, created by G.\ Garbulsky~\cite{ceder1994}.
The inequality coefficients are mostly integers in the range $\pm 100$ with a few larger values.
\item \bvseven: an extended formulation of the permutahedron based on the  Birkhoff-Von Neumann polytope.
It is described by $p^2$ inequalities and $3p-1$ equations in $p^2+p$ variables and also has $p!$ vertices.
The inequalities are all $0,\pm 1$ valued and the equations have single digit integers. The input matrix
is very sparse and the polytope is
highly degenerate.
\end{itemize}

The new problems are:
\begin{itemize}
\item \kmtwo: the Klee-Minty cube for $d=22$ using the formulation given in Chv{\'a}tal~\cite{Chvatal}.
It is non-degenerate and the input coefficients use large integers.
\item \vffive, \vfnine: two random polytopes used in Fisikopoulos and Pe{\~{n}}aranda~\cite{FP16} chosen from input files
kindly provided by the authors.
\vffive consists of 500 random points on a 6-dimensional sphere centred at the origin of
radius 100, rounded to rationals. \vfnine consists of 900 random points in
a 6-dimensional hypercube with
vertices having coordinates $\pm100$.
\item \mitseven: a correlation polytope related to problem \mitine, created by G.\ Garbulsky~\cite{ceder1994}.
The coefficients are similar to \mitine and it is moderately degenerate.
\item \fqfour: related to the travelling salesman problem for on 5 cities, created by F.\ Quondam
(private communication).
The coefficients are all $0,\pm 1$ valued and it is moderately degenerate.
\item \zfw: $0, \pm1$ polytope based on a sensor network that is extremely degenerate
and has large output size, created by Z.F.\ Wang~\cite{zfw14}. There are three non-zeroes per row.
\item \cpsix: the cut polytope for the complete graph $K_6$ solved in the `reverse' direction: from an H-representation to a V-representation.
The output consists of the 32 cut vectors of $K_6$. It is extremely degenerate,  approaching the lower bound
of 19 vertices implied by (\ref{ubt}) for these parameters.
The coefficients of the variables are $0, \pm 1, \pm 2$.
\end{itemize}

\begin{table}[htbp]
\centering
\begin{threeparttable}
\scalebox{0.8}{
\begin{tabular}[t]{|c||c|c|c|c||c|c||c|c|c|c|}
  \hline
  Name &  \multicolumn{4}{|c||}{Input} & \multicolumn{2}{|c||}{Output} & \multicolumn{4}{|c|}{\lrs}\\
    & H/V & $m$ &$n$&size& V/H      & size & bases   &secs         &depth& degeneracy  \\
  \hline
\cthirty &V&30  &16& 4.7K& 341088   & 73.8M&319770   & 43          & 14    & 1  \\
\cforty  &V&40  &21& 12K & 40060020 & 15.6G&20030010 &10002        & 19    & 1  \\
\kmtwo   &H&44  &23& 4.8K& 4194304  &1.2G  & 4194304 & 200         & 22    & 1  \\
\permten &H&1023&11& 29K & 3628800  &127M  & 3628800 & 2381        & 45    & 1  \\
\hline
\vffive  &V&500 &7 & 98K & 56669    &38M   & 202985  & 188         & 41    & 4  \\
\vfnine  &V&900 &7 & 20K & 55903    & 3.9M & 264385  & 97\tnote{1} & 45    & 5  \\
\mitseven&H&71  &61&9.5K & 3149579  &1.1G  & 57613364& 21920       & 20    & 18 \\
\fqfour  &H&48  &19& 2.1K& 119184   &8.7M  & 7843390 & 275         & 24    & 66 \\
\mitine  &H&729 &9 & 21K & 4862     & 196K & 1375608 & 519         & 101   & 283\\
\bvseven &H&69  &57& 8.1K& 5040     & 867K &84707280 &9040         & 17    & 16807   \\
\zfw     &H& 91 &38&7.1K & 2787415  & 205M &10819289888124\tnote{2}&- &-   & 3881478\\
\cpsix   &H& 368&16& 18K & 32       & 1.6K &4844923002&1774681\tnote{3}&153&  151403843\\
  \hline
\end{tabular}
}
\begin{tablenotes}
\item[1] {\cite{FP16} reports an average of 900 secs
for problems like this on an
Intel i5-2400 (3.1GHz)}.
\item[2]{Computed by \mplrsa v.\ 6.2 in 2144809 seconds using 289 cores.}
\item[3] {Computed by \lrs v.\ 6.0.}
\end{tablenotes}
\caption{{Polytopes tested and \lrs times (\mait):  *=time $>$ 604800 secs}}
\label{polytopes}
\end{threeparttable}
\end{table}

We tested five sequential codes, including four based on the
double description method and one based on pivoting:
\begin{itemize}
\item
\cdd(v.\ 0.77): Double description code developed by K.\ Fukuda~\cite{cdd}.
\item
\normaliz(v.\ 3.1.3): Hybrid parallel double description code developed by the Normaliz project~\cite{norm}.
\item
\porta(v.\ 1.4.1): Double description code developed by T.\ Christof and A.\ Lobel~\cite{porta}.
\item
\ppl(v.\ 1.2): Double description code developed by the Parma Polyhedra Library project~\cite{ppl}.
\item
\lrs(v.\ 6.2): C vertex enumeration code based on reverse search developed by D.\ Avis~\cite{lrs}.
\end{itemize}
\noindent
All codes were downloaded from the websites cited and installed using instructions given therein.
Of these, \lrs and \normaliz offer parallelization. For \normaliz this
occurs automatically if it is run on a shared memory multicore machine.
The number of cores used can be controlled with the -x option, which we used extensively in our tests.
For \lrs two wrappers have been developed:
\begin{itemize}
\item
\plrs(v.\ 6.2): C++ wrapper for \lrs using the Boost library, developed by G.\ Roumanis~\cite{AR13}.
It runs on a single shared memory multicore machine.
\item
\mplrs(v.\ 6.2): C wrapper for \lrs using the MPI library, developed by the authors.
\end{itemize}
\noindent

All of the above codes compute in exact integer arithmetic and with the
exception of \porta, are compiled with the GMP library for this purpose.
However \normaliz also uses hybrid arithmetic, giving a very large speedup for
certain inputs as described in the next section.
\porta can also be run in either fixed or extended precision.
Finally, \lrs is also available in a fixed precision 64-bit version, \lrsa, which does no overflow checking.
In general this can give unpredictable results that need independent verification.
In practice, for cases when there is no arithmetic overflow,
\lrsa runs about 4--6 times faster than \lrs (see Computational Results on
the \lrs home page~\cite{lrs}).
The parallel version of \lrsa
(\mplrsa) was used to compute the number of cobases for \zfw, taking roughly 25 days on 289 cores.

\subsection{Sequential Results}
\label{subsec:sequential}

Table \ref{tab:single} contains the results obtained by running the five sequential codes
on the problems described in Table \ref{polytopes}.
Except for \cpsix, the time limit set was one week (604,800 seconds).
Both \normaliz and \porta rejected the problem \vffive due to rational numbers
in the input, as indicated by the letter ``r'' in the table.
For each polytope the first line lists the time in seconds and the second line the space used in megabytes. A hyphen indicates that the space usage was not recorded.
These data were obtained by using the utility \texttt{/usr/bin/time~-a}.

\begin{table}[htb]
\begin{threeparttable}
\centering
\scalebox{0.9}{
\begin{tabular}{|c||c||c||c||c|c||c|}
 \hline

Name     &{\lrs} & \cdd  & \ppl  &\multicolumn{2}{|c||}{\norm}&\porta \\ 
         &secs/MB&secs/MB&secs/MB& (H) secs/MB & (G) secs/MB  &secs/MB\\
\hline
\cthirty &  43   & 2734  &  844  &      27     &      29      &   **  \\
         &  6    & 1701  & 1733  &    2193     &     2193     &   -   \\
 \hline
\cforty  & 10002 &  **   &  *    &    3695     &     4813     &   **  \\
         &   12  &  -    &  -    &    328819   &    328846    &   -   \\
 \hline
\kmtwo   &  200  & 156037&374160 &    1898     &     1776     &   **  \\
         &   6   & 22028 & 31761 &    75189    &     75202    &    -  \\
 \hline
\permten &  2381 &   *   &  *    &    1247     &     14636    &    *  \\
         &   99  & 4904  &  -    &    26018    &     31971    &    -  \\
 \hline
 \hline
\vffive  &  188  &  4385 &  321  &      r      &       r      &    r  \\
         &  69   &  240  &  287  &      -      &       -      &    -  \\
 \hline
\vfnine  &  97   &  3443 &  1004 &      96     &      131     &   **  \\
         &  72   &  148  &  173  &      218    &      194     &    -  \\
 \hline
\mitseven& 21920 &   *   & 91409 &     7901    &     10333    & 109953\\
         &  21   &   -   & 40538 &    115983   &    146226    & 35939 \\
 \hline
\fqfour  &  275  &  438  &  628  &      39     &      287     &  5183 \\
         &   6   &  527  &  983  &     1427    &      1820    &  1141 \\
 \hline
\mitine  &  519  &  440  & 21944 &     203     &      2364    & 47697 \\
         &  71   &  43   &   915 &     337     &       720    & 5623  \\
 \hline
\bvseven &  9040 & 4038  &  477  &     165     &       322    &  296  \\
         &  12   & 1351  & 2073  &     333     &       748    & 457   \\
 \hline
\zfw     &  *    &   *   &   *   &   176606    &       *      & 31120 \\
         &  -    &   -   &   -   &    64668    &       -      & 15944 \\
\hline
\cpsix   & 1774681\tnote{1}&1463829&$>$6570000\tnote{1} &     142329   &1518785\tnote{1} &  $>$4925580 \\
         &  62   &   -   & 13236 &    166226   &       -      &   -   \\
\hline
\end{tabular}
}
\begin{tablenotes}
\item[1]{Codes used were \lrs v.\ 6.0, \ppl v.\ 1.1 and \norm v.\ 3.0.0.} respectively.
\end{tablenotes}
\caption{{Sequential times (\mait):  *=time $>$ 604800 secs  **=abnormal termination}} 
\label{tab:single}
\end{threeparttable}
\end{table}

\cdd, \lrs, and \ppl were used with no parameters.
\norm performs many additional functions, but
was set to perform only vertex/facet enumeration.
By default, it begins with 64-bit integers
and switches to GMP arithmetic (used by all others except \porta) in case
of overflow. In this case, all work done with 64-bit arithmetic is discarded.
Using option -B, \norm will do all computations using GMP.
In Table \ref{tab:single}, we
give times for the default hybrid (H) and for GMP-only (G) arithmetic.
\porta supports arithmetic using 64-bit integers or,
with the -l flag,
its own extended precision arithmetic package.
It terminates if overflow occurs.
We tested both on each problem and found the extended precision
option outperformed the 64-bit option in all cases, so give only the former in the table.

There can be significant variations in the time
of a run.  One cause is dynamic overclocking, where the speed of cores may be increased
by 25\%--30\% when other cores are idle.
Other factors are excessive memory and disk usage,
perhaps by other processes.
Due to the one week time limit and
long \cpsix runs it was not practical to do all runs on otherwise idle machines.
Table \ref{tab:single} should be taken as indicative only.
The two codes which allow parallelization were primarily run on idle
machines
as they are used as benchmarks in Section \ref{parallel}.
In particular, all runs of \lrs (except \zfw and \cpsix due to their length) and all
runs of \norm were
done on otherwise idle machines. These times would
probably increase
by at least the above amounts on a busy machine.
Some times for \cpsix used earlier versions of the
codes, see the table footnotes.
These were not rerun with new versions due to the long running times.

\subsection{Parallel Results}
\label{parallel}

We now give results comparing the three parallel codes using default settings.
For \mplrs and \plrs these are (see User's guide at~\cite{lrs} for details):
\begin{itemize}
\item \plrs: -id 4
\item \mplrs: -id 2, -lmin 3 -maxc 50 -scale 100
\end{itemize}
\noindent
Our main measures of performance are the elapsed time taken and
the {\em efficiency} defined as:
\begin{equation}
\mathrm{efficiency ~=~ \frac{single~core~running~time}{number~of~cores * multicore~running~time}}\,.
\label{eff}
\end{equation}
Multiplying efficiency by the number of cores gives the speedup.
Speedups that scale linearly with the number of cores give constant efficiency.

\begin{table}[h!]
\centering
\scalebox{0.74}{
\begin{tabular}[ht]{|c||c|c|c||c|c|c||c|c|c||c|c|c|}
  \hline
  Name& \multicolumn{3}{|c||}{4 cores} & \multicolumn{3}{|c||}{8 cores}
& \multicolumn{3}{|c||}{12 cores} & \multicolumn{3}{|c|}{16 cores}
 \\
    &\multicolumn{3}{|c||}{secs/efficiency} & \multicolumn{3}{|c||}{secs/efficiency}
& \multicolumn{3}{|c||}{secs/efficiency} & \multicolumn{3}{|c|}{secs/efficiency}
 \\
         &\mplrs&\plrs &\norm  &\mplrs & \plrs&\norm &\mplrs &\plrs & \norm &\mplrs &\plrs &\norm  \\
  \hline
\cforty   &5979 & 3628 & 2475  &  2023 & 2564 & 2131 & 1219  & 2237 & 2048  &  873  & 2066 & 2256  \\
          & .42 & .69  &  .37  &  .62  &  .49 & .22  &  .69  & .37  &  .15  & .72   & .30  & .10   \\
  \hline
\kmtwo    & 190 & 95   &  823  &   65  &  84  &  551 &  39   &  85  &  461  &   28  &  82  & 425   \\
          & .26 &  .53 &  .58  &  .38  &  .30 & .43  &  .43  & .20  &  .34  &  .45  &  .15 &  .28  \\
  \hline
\permten  &1422 &  709 & 1232  &  481  &  445 & 1100 &  292  &  367 &  1067 &  215  & 320  & 1061  \\
          & .42 &  .84 &  .25  &  .62  &  .67 &  .14 &  .68  & .54  &  .10  &  .69  & .47  & .07   \\
  \hline
  \hline
\vffive   & 92  &  46  &   r   &   36  &  27  &   r  &  22   &  19  &   r   &   17  &  19  &  r    \\
          & .51 & 1.02 &   -   &  .65  &  .87 &   -  &  .71  & .82  &   -   &  .69  & .62  &  -    \\
  \hline
\vfnine   &  51 &  26  &  36   &   20  &  16  &   22 &   11  &  13  &   28  &   9   &   11 &  100  \\
          & .48 & .93  & .67   &  .61  & .76  &  .55 &  .73  & .62  &  .29  & .67   &  .55 &  .06  \\
  \hline
\mitseven &11386& 6479 & 2452  & 3983  & 3320 & 1360 &  2390 & 2254 &  973  & 1709  & 1724  & 798  \\
          & .48 &  .85 &  .81  &  .69  &  .83 & .73  &  .76  & .81  &  .68  & .80   &  .79  & .62  \\
  \hline
\fqfour   & 146 &  70  &  15   &   49  &  37  &  10  &   30  &  27  &   8   &   21  &   21  &  9   \\
          & .47 & .98  &  .65  &  .70  & .93  &  .49 &  .76  & .85  &  .41  &  .82  &  .82  & .27  \\
  \hline
\mitine   & 293 &  152 &  89   &   99  &  89  &  51  &   61  &  68  &   39  &  44   &   57  &  39  \\
          & .44 & .85  &  .57  &  .66  &  .73 & .50  &  .71  & .64  &  .43  &  .74  &  .57  & .33  \\
  \hline
\bvseven  &5219 & 2399 &  47   &  1739 & 1213 &  26  & 1045  &  818 &  18   &  747  &  624  &  14  \\
          & .43 & .94  & .88   &  .65  &  .93 & .79  &  .72  &  .92 & .76   & .76   & .91   & .74  \\
  \hline
\zfw      &  *  &  *   & 49246 &   *   &   *  &24057 &   *   &   *  & 16686 &   *   &   *   & 13160\\
          &  -  &  -   &  .90  &   -   &   -  & .92  &   -   &   -  & .88   &   -   &   -   & .84  \\
\hline
\cpsix   &968550&486667& 43360 &331235 &268066&24520 &199501 &201792& 18016 &143006 &169352 & 15301\\
         & .46  &  .91 &  .82  &  .67  &  .83 & .73  &  .74  &  .73 &  .66  & .78   & .65   & .58  \\
\hline
\end{tabular}
}
\caption{Small scale parallelization (\mait):  *=time $>$ 604800 secs, **=abnormal termination}
\label{tab:su12}
\end{table}

Table \ref{tab:su12} gives results for low scale parallelization
using \mait. We omit \cthirty as it runs in under a minute
using a single processor with either \lrs or \norm.
We observe that for \plrs and \norm the efficiency goes down as the number of cores
increases as is typical for parallel algorithms. The efficiency of \mplrs, however, 
goes up.
This is due to the fact we assign one core each to the master and consumer which 
continually
monitor the remaining worker cores which run \lrs. Therefore with 16 cores there are
14 workers which is 7 times as many workers as when 4 cores are used; hence
the improved efficiency. We discuss this further in Section \ref{analysis}.

For \cpsix, the \lrs times in Tables \ref{polytopes}--\ref{tab:single}
were obtained using v.\ 6.0 which has a smaller backtrack cache size than v.\ 6.2.
Hence the \mplrs and \plrs speedups against \lrs for \cpsix
in Table \ref{tab:su12} are probably somewhat larger than they would be against \lrs v.\ 6.2.
With 4 cores available, \plrs
usually outperforms \mplrs, they give similar performances with 8 cores,
and \mplrs
is usually faster with 12 or more cores.
With 16 cores \mplrs gave quite consistent performance with efficiency in the range .67 to .82,
with the exception of \kmtwo with efficiency .45.
The efficiencies obtained by
\plrs and \norm show a much higher variance, in the range .15 to .91 and .06 to .84
respectively.

Table \ref{tab:su64} contains results for medium scale parallelization on
the 64-core shared memory machine \maief.
We omit from the table the five problems that \mplrs could solve in under a minute with 16 cores.
Note that these processors are considerably
slower than \mait on a per-core basis as can be seen by comparing the single processor times in Tables \ref{tab:single} and \ref{tab:su64}.
The running time for \lrs on \cpsix was estimated by scaling the time for a partial run,
making use of the fact that \lrs runs in time proportional to the number of bases computed.
In this case the partial run produced 1807251355 bases in 1285320 seconds. So we
scaled up this running time using the known total number of bases
given in Table \ref{polytopes}.

\begin{table}[htbp]
\centering
\begin{threeparttable}
\scalebox{0.73}{
\begin{tabular}[ht]{|c||c|c||c|c|c||c|c|c||c|c|c||c|c|c|c|c|}
  \hline
  Name& \multicolumn{2}{|c||}{1 core} &\multicolumn{3}{|c||}{16 cores} & \multicolumn{3}{|c||}{32 cores}
& \multicolumn{5}{|c|}{64 cores}
 \\
    &\multicolumn{2}{|c||}{secs/efficiency}  &\multicolumn{3}{|c||}{secs/efficiency} & \multicolumn{3}{|c||}{secs/efficiency}
& \multicolumn{3}{|c||}{secs/efficiency} & \multicolumn{2}{|c|}{memory (MB)}
 \\
         & \lrs           & \norm&\mplrs&\plrs &\norm&\mplrs&\plrs &\norm&\mplrs&\plrs&\norm&\plrs&\norm\\
  \hline
\cforty  & 15039          & 5464 & 1453 & 3711 & 3647&  782 & 3607 & 4421& 466 & 3561 & 4593& 154&100839\\
         & 1              &   1  &  .65 & .25  & .09 &  .60 &  .13 & .04 & .50 &  .07 & .02 &    &      \\
\hline
\permten & 3741           & 2420 &  371 &  543 & 1645&  207 &  556 &1638 & 140 &  509 & 1930& 771& 8063 \\
         & 1              &   1  &  .63 & .43  & .09 &  .56 &  .21 & .05 & .42 &  .11 & .02 &    &      \\
\hline
\hline
\mitseven& 35426          & 17448& 2965 & 3367 &2831 & 1592 & 1806 & 1694&1040 & 1368 & 1163& 385& 29689\\
         & 1              &   1  &  .75 & .66  & .39 &  .70 &  .61 & .32 & .53 &  .40 & .23 &    &      \\
\hline
\bvseven & 14340          & 333  & 1271 & 1188 &  44 &  683 &  612 &  30 & 460 &  434 &  22 & 149& 139  \\
         & 1              &   1  &  .71 & .75  & .47 &  .66 &  .73 & .35 & .49 &  .52 & .24 &    &      \\
\hline
\zfw     & *              &289813&   *  &   *  &21604&   *  &   *  &12600&  *  &   *  & 6768&  * & 21829\\
         & -              &   1  &   -  &   -  & .84 &   -  &   -  & .72 &  -  &   -  & .67 &    &      \\
\hline
\cpsix   &3445717\tnote{1}&191586&312264&367249&26517&183161&260200&18740&90296&200801&15758&1218&43270 \\
         & 1              &   1  & .69  & .59  & .45 &  .59 &  .41 & .32 & .60 &  .27 & .19 &    &      \\
\hline

\end{tabular}
}
\begin{tablenotes}
\item[1]{Estimate based on scaling a partial run on the same machine.}
\end{tablenotes}
\caption{Medium scale parallelization (\maief):  *=time $>$ 604800 secs, **=abnormal termination}
\label{tab:su64}
\end{threeparttable}
\end{table}

With 64 cores, in terms of efficiency,
\mplrs again gave a very consistent performance with efficiencies ranging from .42 to .60.
This compares to .07 to .52 for \plrs and .02 to .67 for \norm.
We give memory usage for the 64 core runs for \plrs and \norm. Memory usage by \mplrs
is not directly measurable by the \texttt{time} command mentioned above, but is comparable to
\plrs. On problem \cpsix, with 64 cores \norm is nearly 6 times
faster than \mplrs but this is due to the arithmetic package. On a similar run
using GMP arithmetic, \norm took 182236 seconds which is twice as long as \mplrs.

For this scale of parallelization some limited computational results for \prs were given in \cite{BMFN99}.
They report in detail on only one problem which has an input size of
$m=134$ and $n=11$ obtaining efficiencies of .94, .35 and .26, respectively, when using 
10, 100 and 150 processors on a Paragon MP computer. Their problem solves in under a minute
with the current version of \lrs so no direct comparison of efficiency with \mplrs is possible. The authors also
report solving three 
problems for the first time including
\mitseven, which completed in 4.5 days using 64 processors on a Cenju-3.
They estimated the single processor running time for \mitseven to be 
130 days on a DEC AXP. This machine has a very different processor and architecture making
it hard to meaningfully estimate the efficiency of the Cenju-3 run.
\begin{table}[htbp]
\centering
\scalebox{0.9}{
\begin{tabular}[ht]{|c||c|c|c|c||c|c|}
  \hline
 Name& \multicolumn{6}{|c|}{\mplrs~~~~~secs/efficiency}  \\
  &96 cores&128 cores&160 cores&192 cores&256 cores& 312 cores\\
\hline
\cforty  &  329  &  247  &   203 &  179  &  134  & 129   \\
         &  .48  &  .48  &  .46  &  .44  & (.44) & (.37) \\
\hline
\permten &  115  &  94   &   85  &   96  &   64  &  61   \\
         &  .34  &  .31  &  .28  &  .20  & (.23) & (.20) \\
\hline
\hline
\mitseven&  686  &  516  &  412  &  350  &  231  &  205  \\
         &  .54  &  .54  &  .54  &  .53  & (.60) & (.55) \\
\hline
\bvseven &  302  &  229  &  184  &  158  &  98   &  88   \\
         &  .49  &  .49  &  .49  &  .47  & (.57) & (.52) \\
\hline
\cpsix   & 56700 & 43455 & 34457 & 28634 & 18657 & 15995 \\
         &  .63  &  .62  &  .63  &  .63  & (.72) & (.69) \\
\hline
\end{tabular}
}
\caption{Large scale parallelization (\maicl cluster)}
\label{tab:su96}
\end{table}

Table \ref{tab:su96} contains results for large scale parallelization
on the 312-core \maicl cluster of 9 nodes described in Section \ref{subsec:expsetup}.
Only \mplrs can use all cores in this heterogeneous environment.
The first 5 columns used only the \maitt group of five nodes which all use the same
processor. The efficiencies are therefore directly comparable and Table \ref{tab:su96} is an extension of Table \ref{tab:su64}.
In the final two columns the machines were scheduled in the order given in Section \ref{subsec:expsetup}.
Since the processors have different clock speeds
we include the efficiency in parentheses as it is only a rough estimate.

Finally, Table \ref{tab:su1200} shows results for very large scale parallelization
on the \tsubame supercomputer at
the Tokyo Institute of Technology. We ran tests on the four hardest problems for \mplrs.

\begin{table}[htbp]
\centering
\begin{threeparttable}
\scalebox{0.9}{
\begin{tabular}[ht]{|c||c|c|c|c|c|}
  \hline
  Name    &            \multicolumn{5}{|c|}{\mplrs }         \\

          & 1 core  &300 cores&600 cores&900 cores&1200 cores\\
  \hline
\cforty   &  17755  &   89    &    49   &   43    &    44    \\
          &      1  &   .66   &   .60   &  .46    &   .34    \\
  \hline
\hline
\mitseven &  36198  &   147   &    80   &   63    &   49     \\
          &      1  &   .82   &   .75   &  .64    &  .62     \\
\hline
\bvseven  &  10594  &    48   &    27   &   27    &   29     \\
          &      1  &   .73   &   .65   &  .44    &  .30     \\
  \hline
\cpsix    &2400648\tnote{1}& 9640& 4887 &  3278   &  2570    \\
          &      1  & .83     &  .82    &  .81    &  .78     \\
  \hline
\end{tabular}
}
\begin{tablenotes}
\item[1]{Estimate based on scaling a partial run on the same machine.}
\end{tablenotes}
\caption{Very large scale parallelization: secs/efficiency}
\label{tab:su1200}
\end{threeparttable}
\end{table}

The hardest problem solved was \cpsix, 
the 6 point cut polytope solved in the reverse direction, which is
extremely degenerate. Its more than 4.8 billion bases span just 32
vertices! Normally such polytopes would be out of reach for pivoting
algorithms.
We observe 
near linear speedup between
300 and 1200 cores. 
Solving in the `reverse' direction is useful for checking
the accuracy of a solution, and is usually extremely time consuming. For example,
converting the V-representation of \cpsix to an H-representation takes less than
2 seconds using any of the three single core codes.

\subsection{Analysis of Results}
\label{analysis}

\begin{figure}[htb]
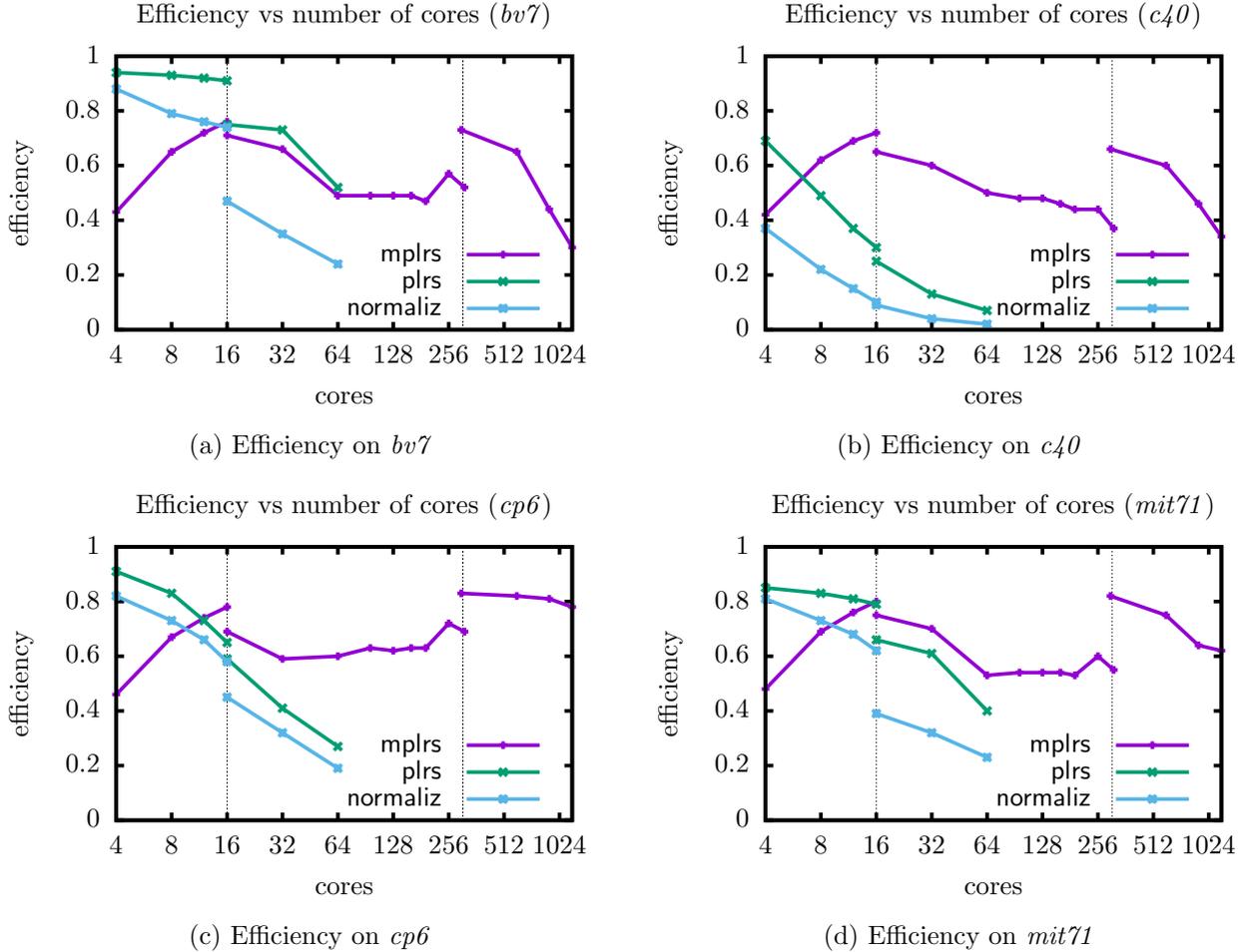

\centering
\begin{subfigure}[b]{0.49\textwidth}
 \centering
 \resizebox{\textwidth}{!}{ \input{plots/log-efficiency-bv7.tex} }
 \caption{Efficiency on \bvseven}
 \label{subfig:bv7logefficiency}
\end{subfigure}
\begin{subfigure}[b]{0.49\textwidth}
 \centering
 \resizebox{\textwidth}{!}{ \input{plots/log-efficiency-c40.tex} }
 \caption{Efficiency on \cforty}
 \label{subfig:c40logefficiency}
\end{subfigure}
\begin{subfigure}[b]{0.49\textwidth}
 \centering
 \resizebox{\textwidth}{!}{ \input{plots/log-efficiency-cp6.tex} }
 \caption{Efficiency on \cpsix}
 \label{subfig:cp6logefficiency}
\end{subfigure}
\begin{subfigure}[b]{0.49\textwidth}
 \centering
 \resizebox{\textwidth}{!}{ \input{plots/log-efficiency-mit71.tex} }
 \caption{Efficiency on \mitseven}
 \label{subfig:mit71logefficiency}
\end{subfigure}
\caption{Efficiency vs number of cores (data from Tables \ref{tab:su12}--\ref{tab:su1200} )}
\label{fig:logefficiency}
\end{figure}

In Figure \ref{fig:logefficiency} we plot the efficiencies of the three
parallel codes on the four hardest problems that they could all solve,
using a logarithmic scale for the horizontal axis.
Each figure is divided into three parts by two vertical lines.
The left part corresponds to data from Table \ref{tab:su12},
the centre part to data from Tables \ref{tab:su64}--\ref{tab:su96} and the right part to
data from Table \ref{tab:su1200}.
Recall that speedup is the product of efficiency times the number of cores, and that
a horizontal line in the figure corresponds to speedups that scale linearly with the
number of cores.
Overall near linear speedup is observed for \mplrs throughout the
range until about 500 cores and, in two cases,
until 1200 cores.
The efficiencies for \plrs and \norm generally decrease monotonically to
64 cores, the limit of our shared memory hardware.

The \mplrs plots have more or less the same shape. In the left section the
efficiency increases. This is due to the fact that one core is used as the
master process and one as the collection process. Therefore there
are 2 \lrs workers
when 4 cores are available which rises to 14 workers with 16 cores, a 7 fold increase.
There is a small drop in efficiency at 16 cores as \maief replaces the more
powerful \mait. A similar drop is observable for \plrs and \norm.
A small increase
in efficiency is observed at 256 cores as
\mait is used in the cluster and hosts the master/collector processes.
Finally a jump occurs at 300 cores as
\tsubame replaces the \maicl cluster and then efficiency decreases.

A decrease in efficiency indicates that overhead has increased. 
The two causes of overhead in \plrs discussed in Section~\ref{sec:plrs}
remain in \mplrs.
One cause is the cost, for each job taken from $L$, of pivoting to
the LP dictionary corresponding to its restart basis.
This is borne by each worker
as it receives a new job from the list $L$. This cost is directly proportional to the
length of the job list, which is typically longer in \mplrs than in \plrs.
However, this overhead is shared among all workers and so the cost is
mitigated. 
The amount of overhead for each job depends on the number of pivots to be made and
on the difficulty of an individual pivot. It is therefore highly problem dependent and
this is one reason why the efficiency varies from problem to problem.

The second cause of overhead is that processors are idle when $L$ becomes empty. In Section~\ref{sec:plrs}
we saw that this was a major problem with \plrs as this overhead {\em increases} as more
and more processors become idle when $L$ is empty. This overhead has been largely eliminated
in \mplrs by our budgeting and scaling strategy, as $L$ rarely becomes empty.
This was illustrated in Figure~\ref{fig:hist_sample}(b). 
A third cause of overhead in \mplrs are the
master and the consumer processes, as mentioned above. This overhead was not apparent in \plrs.
It dissipates, however, as the number of cores increased
as we see in Figure~\ref{fig:logefficiency}.

There is additional overhead and bottlenecks in \mplrs due to 
communication between nodes.  For instances such as \cforty that have large
output sizes, the workers can saturate the interconnect.  In 
Table~\ref{tab:su96}, the times for \cforty slightly beat the time needed
to transfer the output over the gigabit ethernet interconnect (which is
possible because some of the workers are local to the collector and so
some of the output does not need to be transferred).  One could transfer
the output in a more compact form, but this would involve additional
modifications to the underlying \lrs code.

The latency involved in communications is also an issue, since we pay
this cost each time we send a job to a worker.  This is especially costly
on small jobs, which can be very common 
(cf. Figure~\ref{fig:hist_sample}(d)).  The lower latency
of the Tsubame interconnect is likely responsible for the jump in efficiency
that we see at 300 cores in Figure~\ref{fig:logefficiency} (and also
the higher bandwidth in the case of \cforty).

Ideally, an algorithm that scales perfectly would have an efficiency of 1
for any number of cores.
However our present hardware does not seem able to achieve this due to a combination of factors.
As a test, we ran
multiple copies of \lrs in parallel and computed the efficiency,
compared to the same number of sequential single runs, using (\ref{eff}).
Specifically, using the problem \mitine
we ran, respectively, 16, 32 and 64 copies of \lrs in parallel on
the 64-core
\maief.
The time
of a single \lrs run on this machine is 892 seconds
and the times of the parallel runs were, respectively, 958, 1060 and 1465 seconds.
So the efficiencies obtained
were respectively .93, .84 and .61.
One possible cause for this is that dynamic overclocking (mentioned in
Section~\ref{subsec:sequential})
limits the maximum efficiency obtainable by the parallel codes.
However, leaving some cores idle in order to obtain higher frequencies
on working cores is a technique worth consideration and so we did not disable
dynamic overclocking.

\begin{figure}[htb]
\centering
\begin{subfigure}[b]{0.47\textwidth}
 \centering
 \resizebox{\textwidth}{!}{
   \input{plots/maxc1.tex}
 }
 \caption{$\maxcobases=1$~:~1132 secs, $|L|=\mbox{2,157,153}$}
\end{subfigure}
\begin{subfigure}[b]{0.47\textwidth}
 \centering
 \resizebox{\textwidth}{!}{
   \input{plots/maxc10.tex}
 }
 \caption{$\maxcobases=10$~:~604 secs, $|L|=\mbox{417,272}$}
\end{subfigure}
\begin{subfigure}[b]{0.47\textwidth}
 \centering
 \resizebox{\textwidth}{!}{
   \input{plots/maxc100.tex}
 }
 \caption{$\maxcobases=100$~:~501 secs, $|L|=\mbox{69,422}$}
\end{subfigure}
\begin{subfigure}[b]{0.47\textwidth}
 \centering
 \resizebox{\textwidth}{!}{
   \input{plots/maxc1000.tex}
 }
 \caption{$\maxcobases=1000$~:~516 secs, $|L|=\mbox{30,088}$}
\end{subfigure}

\caption{The effect of varying the budget parameter $\maxcobases$ 
         (\maief,\mainew)~:~\mitseven,~128 cores,~$\myscale=100$} 
\label{fig:budget}
\end{figure}

Finally we address the sensitivity of the performance of \mplrs to the two
main parameters, \maxcobases and \myscale.
Here the news is encouraging: the running time is quite stable
over a wide range of values for the problems we have tested.
Figure \ref{fig:budget}
shows the job list evolution and running times for \mitseven using 128 cores
on \maief and \mainew with $\maxcobases=1,10,100,1000$.
Recall that  Figure \ref{fig:hist_sample}(b) contains
the histogram for the default setting of $\maxcobases=50$, where
a total of 120,556 jobs were created
and the running time was 516 seconds.
We observe that, apart from the extreme value $\maxcobases=1$, the running time is quite stable in
the range of 500--600 seconds, for very different budgets.
Note that the number of jobs produced does vary
a lot. With $\maxcobases=1000$ the job queue becomes dangerously near empty at roughly 110
and 200 seconds and for the last 40 seconds.
The other three job queue plots show similar behaviour and $\maxcobases=100$ wins the race
since it generates the fewest extra jobs.

\begin{figure}[htb]
\centering
\begin{subfigure}[b]{0.47\textwidth}
 \centering
 \resizebox{\textwidth}{!}{
   \input{plots/scale1.tex}
 }
 \caption{$\myscale=1$~:~1482 secs, $|L|=\mbox{3,315,660}$}
\end{subfigure}
\begin{subfigure}[b]{0.47\textwidth}
 \centering
 \resizebox{\textwidth}{!}{
   \input{plots/scale10.tex}
 }
 \caption{$\myscale=10$~:~685 secs, $|L|=\mbox{683,870}$}
\end{subfigure}
\begin{subfigure}[b]{0.47\textwidth}
 \centering
 \resizebox{\textwidth}{!}{
   \input{plots/scale1000.tex}
 }
 \caption{$\myscale=1000$~:~528 secs, $|L|=\mbox{32,721}$}
\end{subfigure}
\begin{subfigure}[b]{0.47\textwidth}
 \centering
 \resizebox{\textwidth}{!}{
  \input{plots/scale10000.tex}
 }
 \caption{$\myscale=10000$~:~872 secs, $|L|=\mbox{54,555}$}
\end{subfigure}

\caption{The effect of varying the \myscale parameter (\maief,\mainew)~:~\mitseven,~128 cores,~$\maxcobases=50$}
\label{fig:scale}
\end{figure}

Figure \ref{fig:scale} shows the job list evolution and running time with
$\maxcobases=50$
and varying $\myscale=1,10,1000,10000$. Recall 
Figure~\ref{fig:hist_sample}(b) contains the plot for $\myscale=100$.
With a $\myscale=1$ too many jobs are produced, slowing the running time by
nearly a factor of 3 compared to the default settings.
With $\myscale=1000$ we notice that even though the job queue becomes 
empty roughly 50 seconds before
the end of the run the total running time is nearly the same as with default settings. 
The situation is much worse with $\myscale=10000$ as the job list is essentially
empty for almost half of the run.
We see that the number of jobs produced drops rapidly
as the scale is increased up to 1000 but then rises for a scale of 10000. 
This is due to the fact the budget gets reset back to $\maxcobases=50$ 
whenever the job list becomes nearly empty,
which happens frequently in this case.

It would be nice to get a formal relationship between job list size and the budget.
This is likely to be very difficult for the vertex enumeration problem due to vast
differences in search tree shapes. However such results are possible for random
search trees.
In recent work Avis and Devroye \cite{AD17a} analyzed this relationship for 
very large randomly generated Galton-Watson trees.
They showed that, in probability,
the job list size declines as the square root of the increase in budget.

\section{Conclusions}
\label{conclusions}
It is natural to ask what is the limit of the scalability of the current \mplrs?
Very preliminary experiments with \tsubame using up to 2400 cores indicate that this limit may be 
at about 1200 cores. Although budgeting seemed to produce nicely scaled job queue sizes,
there was a limit to the ability of the single producer (and consumer) to keep up with the workers.
While small modifications can perhaps push this limit somewhat further,
this indicates that a higher level `depot' system may be required, where each depot
receives a part of the job queue and acts as a producer with a subset of the available cores.
This could also help avoid overhead related to the interconnect latency, since
many jobs would be available locally and even remote jobs would be transferred 
in blocks.
Similarly the output may need to be collected by several consumers, especially when it is 
extremely large as in \cforty and \mitseven. These are topics for future research.

Finally one may ask if the parallelization method used here could be used
to obtain similar results for other tree search applications.
Indeed we believe it can. In ongoing work \cite{AJ16a} we have prepared a generic framework
called \mts that can be used to parallelize legacy reverse search enumeration codes.
The results presented there for two other reverse search applications
give comparable speedups to the ones we obtained for \mplrs.
We are also extending the range of possible applications by allowing
in \mts a certain amount of shared
information between workers. This allows the possibility of trying this approach
on branch and bound algorithms, game trees, satisfiability solvers, and the like.

\subsubsection*{Acknowledgements}
We thank Kazuki Yoshizoe for
helpful discussions concerning the MPI library
which improved \mplrs' performance.

\bibliographystyle{spmpsci}
\bibliography{paper}

\end{document}